\documentclass[aps,prl,amssymb,superscriptaddress,twocolumn]{revtex4-1}
\usepackage[pdftex]{graphicx}
\usepackage{bm}
\usepackage{amsmath}
\usepackage{ascmac}
\usepackage{color}
\usepackage{setspace}
\usepackage{amssymb}
\usepackage{latexsym}
\usepackage{mathtools}
\usepackage{lipsum}
\usepackage{braket}
\usepackage[colorlinks,linkcolor=blue,citecolor=blue]{hyperref}

\newtheorem{prop}{Proposition}

\newcommand{\sq}{\qquad $\blacksquare$}

\begin{document}

\title{Dynamical Scaling of Surface Roughness and Entanglement Entropy in Disordered Fermion Models} 

\author{Kazuya Fujimoto}
\affiliation{Institute for Advanced Research, Nagoya University, Nagoya 464-8601, Japan}
\affiliation{Department of Applied Physics, Nagoya University, Nagoya 464-8603, Japan}

\author{Ryusuke Hamazaki}
\affiliation{Nonequilibrium Quantum Statistical Mechanics RIKEN Hakubi Research Team, RIKEN Cluster for Pioneering Research (CPR), RIKEN iTHEMS, Wako, Saitama 351-0198, Japan}

\author{Yuki Kawaguchi}
\affiliation{Department of Applied Physics, Nagoya University, Nagoya 464-8603, Japan}

\date{\today}

\begin{abstract}
Localization is one of the most fundamental interference phenomena caused by randomness, and its universal aspects have been extensively explored from the perspective of one-parameter scaling mainly for $static~properties$. 
We numerically study dynamics of fermions on disordered one-dimensional potentials exhibiting localization and find $dynamical$ one-parameter scaling for surface roughness, which represents particle-number fluctuations at a given lengthscale, and for entanglement entropy when the system is in delocalized phases. 
This dynamical scaling corresponds to the Family-Vicsek scaling originally developed in classical surface growth, and the associated scaling exponents depend on the type of disorder. Notably, we find that partially localized states in the delocalized phase of the random-dimer model lead to anomalous scaling, where destructive interference unique to quantum systems leads to exponents unknown for classical systems and clean systems. 
\end{abstract}

\maketitle

{\it Introduction.}
The Anderson localization \cite{ori_AL} is a unique phenomenon arising from destructive interference in disordered systems.
It has attracted a lot of attention in, e.g., solid-state physics, quantum optics, and classical mechanics \cite{AL_Review1, AL_Review2, AL_Review3, AL_Review4}, and has been observed in various experimental setups \cite{AL_ele1,AL_ele2, AL_gas1,AL_gas2,AL_gas3,AL_gas4,AL_gas5,AL_gas6,AL_gas7,AL_Review5,AL_light1,AL_light2,AL_light3,AL_light4,AL_light5,AL_elas1,AL_elas2}. Study of the Anderson localization has significantly been put forward in light of one-parameter scaling \cite{AL_Review2,Scaling1,Scaling2}, where physical quantities are scaled only by a single parameter. The example includes scaling for system-size dependence of conductance and for correlation functions at localization transition points. Despite its importance, such a one-parameter scaling has been focused mainly for static properties. Meanwhile, disorder is known to affect quantum dynamics, such as entanglement dynamics \cite{Sirker1,Sirker2,Sirker3,Sirker4,MBL2,MBL3,MBL4,Nahum1,AL_dynamics4} and transport properties \cite{AAM4,AL_dynamics1,AL_dynamics2,AL_dynamics3,AL_dynamics5,MBL7,MBL8,MBL9,MBL10,MBL11}. 
It is thus intriguing and fundamental to pursue dynamical one-parameter scaling, which can lead to hitherto unknown classification of disordered quantum systems by their nonequilibrium properties.

\begin{figure}[t]
\begin{center}
\includegraphics[keepaspectratio, width=8.6cm,clip]{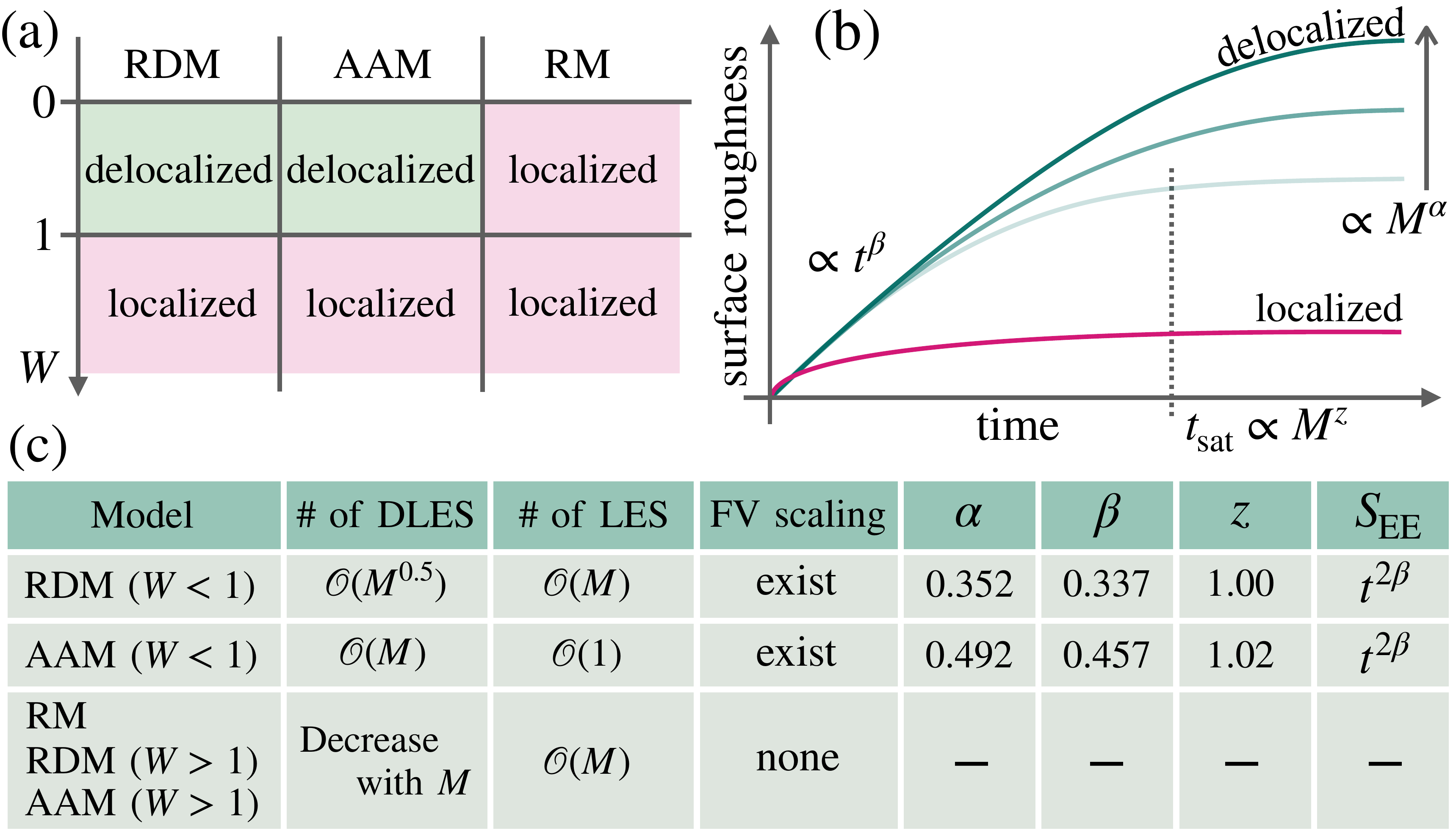}
\caption{
(a) Phase diagram for the random model (RM), the random-dimer model (RDM), and the Aubry-André model (AAM) as a function of the disorder strength $W$. Delocalized phases appear for $W<1$ in the RDM and the AAM. (b) Schematic for surface-roughness growth in the localized and delocalized phases. In the delocalized phase, the growth is characterized by three power exponents $\alpha$, $\beta$, and $z$, which respectively capture system-size $M$ dependence of the saturated surface roughness, power-law growth, and a saturation time $t_{\rm sat}$ of the surface roughness. This dynamical scaling is called the Family-Vicsek (FV) scaling (see Eq.~\eqref{FV1}). The FV scaling does not emerge in the localized phase. (c) Summary of our results, including numbers of delocalized eigenstates (DLESs) and localized eigenstates (LESs) and growth laws of von Neumann entanglement entropy $S_{\rm EE}$. } 
\label{fig1} 
\end{center}
\end{figure}

It has recently been found that dynamical one-parameter scaling, called the Family-Vicsek (FV) scaling, appears in a clean quantum bosonic system \cite{height_op2}. While the FV scaling was originally known in classical surface growth \cite{Vicsek1984,Vicsek1985,barabasi1995fractal}, Ref.~\cite{height_op2} finds the scaling in the quantum system by introducing ``quantum surface-height operator, '' which represents particle-number fluctuations summed over a given lengthscale (see Eq.~\eqref{sh_op}). The standard deviation of this operator, i.e., quantum surface roughness, is found to obey the Edwards-Wilkinson (diffusive) and ballistic scalings. Notably, the surface roughness is experimentally accessible in cold atomic systems using microscopes.

In this Letter, employing the surface roughness in quantum systems, we first show numerical evidence that dynamical one-parameter scaling exists in one-dimensional (1D) non-interacting fermions in a disordered potential. We use the random model (RM), the random-dimer model (RDM) \cite{RDM1}, and the Aubry-André model (AAM) \cite{AAM1}, which exhibit the Anderson localization. The phase diagram of these models is schematically shown in Fig.~\ref{fig1}(a) as a function of disorder strength $W$. Our numerical calculations find that, in the delocalized phases of the RDM and the AAM, the surface roughness obeys the FV scaling characterized by three exponents $\alpha$, $\beta$, and $z$ as schematically shown in Fig.~\ref{fig1}(b). 
Notably, we find anomalous exponents $(\alpha,\beta,\gamma) \simeq (0.352,0.337,1.00)$ in the RDM. 
We argue that the anomalous scaling is caused by numerous localized eigenstates in a delocalized phase, which are unique to quantum disordered systems. Furthermore, we find that the surface roughness is approximately proportional to the square root of the von Neumann entanglement entropy (EE), and our numerical calculation elucidates the FV-type scaling of the EE. Our finding suggests that the surface roughness can be an experimentally friendly measure for the EE. Table in Fig.~\ref{fig1}(c) summarizes our results.

{\it Theoretical models.}
We consider non-interacting $N$-spinless fermions on a 1D lattice with a disordered potential.
Let us denote the annihilation and creation operators on a site $j$ by $\hat{f}_j$ and $\hat{f}_j^{\dagger}~(j=1,\cdots ,M)$, where $M$ is the number of the lattice sites. Throughout this work, $M$ is set to be even.
Then, the Hamiltonian is given by
\begin{eqnarray}
\hat{H} = - J \sum_{j=1}^{M} \left( \hat{f}_{j+1}^{\dagger} \hat{f}_j + \hat{f}_{j}^{\dagger} \hat{f}_{j+1} \right) + \sum_{j=1}^{M} V_j \hat{f}_{j}^{\dagger} \hat{f}_{j}
\end{eqnarray}
with a hopping constant $J>0$ and an on-site potential $V_j$. 

We use three potentials corresponding to the RM, the RDM, and the AAM. 
The RM consists of a random potential with no spatial correlation, where $V_j$ takes $0$ or $V(>0)$ following the probability function $P_{\rm RM}(V_j) = \frac{1}{2} \delta(V_j) + \frac{1}{2} \delta(V_j- V)$.
The potential in the RDM \cite{RDM1} has a spatial correlation such that the probability function is given by 
$P_{\rm RDM}(V_{2j-1},V_{2j}) = \frac{1}{2} \delta(V_{2j})\delta(V_{2j-1}) + \frac{1}{2} \delta(V_{2j}-V) \delta(V_{2j-1}-V)$ with $j=1,2,\cdots, M/2$ \cite{RDM3}.
The AAM has fixed quasi-periodic structure given by $V_j = V \cos (2 \pi \theta j)$ with the irrational number $\theta = (\sqrt{5}-1)/2$ \cite{AAM1,AAM2,AAM3}. 
We assume the periodic boundary condition for the RM and the RDM, and the open boundary condition for the AAM. 
In the RDM and the RM, we take ensemble averages to calculate physical quantities, and the sample number in all the calculations is $\lfloor 44000/M \rfloor$ with the floor function $\lfloor  \cdots \rfloor$.

The strength of the disorder is characterized by the dimensionless constant $W = V/(2J)$. The models have localized or delocalized phases depending on $W$ \cite{RDM1,AAM2} as shown in Fig.~\ref{fig1}(a). In the RM, all the eigenstates are localized in the thermodynamic limit if $W$ is nonzero. 
If the randomness has spatial correlation as for the RDM and the AAM, there exist delocalized phases for $W<1$. 
The RDM has both delocalized eigenstates (DLESs) and localized eigenstates (LESs) in the delocalized phase, but there are no mobility edges (see Sec. I of Supplemental material (SM)~\cite{SM}).

\begin{figure*}[t]
\begin{center}
\includegraphics[keepaspectratio, width=17.8cm,clip]{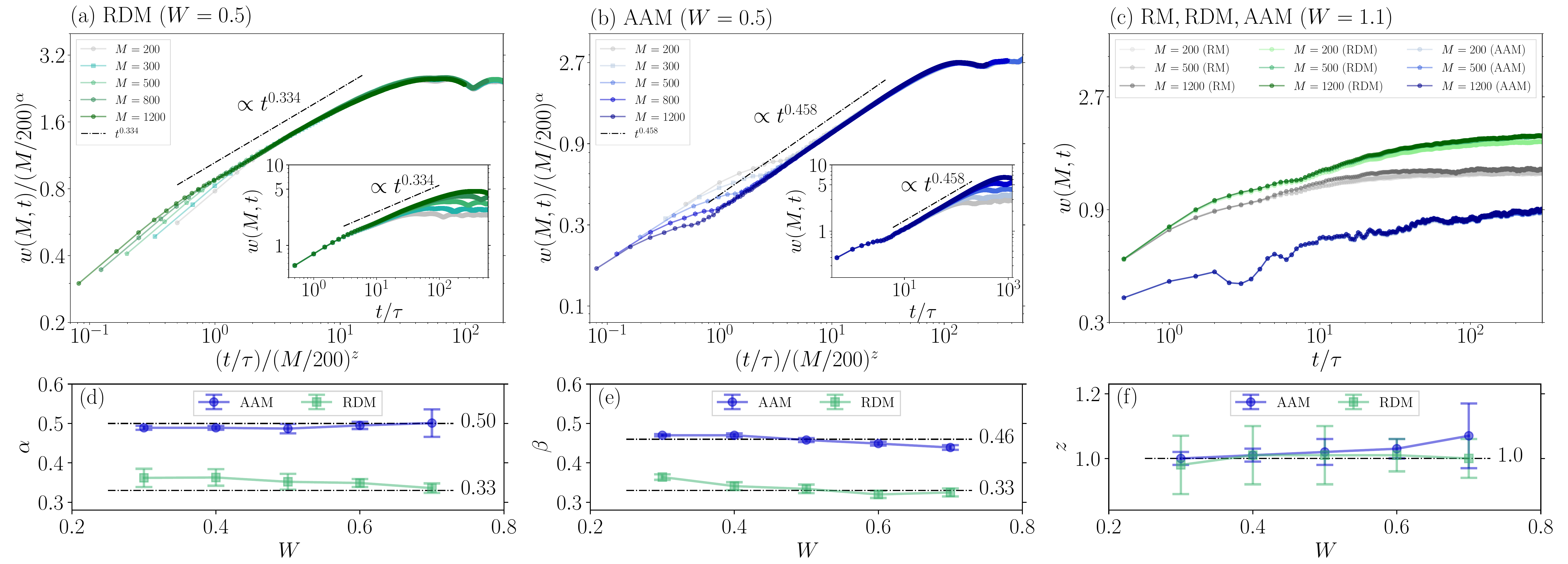}
\caption{Surface-roughness dynamics and FV scaling for delocalized phases of (a) RDM ($W=0.5$) and (b) AAM ($W=0.5$), and for (c) localized phases of RM, RDM, and AAM ($W=1.1$). The time is normalized by $\tau=\hbar/J$. In (a) and (b), the main panels show $w(M,t)$ with $M=200, 300, 500, 800$, and $1200$ with the ordinate and the abscissa normalized by $(M/200)^{\alpha}$ and $(M/200)^z$, and the insets show the corresponding raw data. The delocalized phase shown in (a) and (b) exhibits the clear FV scaling, whereas we find no signature of the FV scaling in the localized phase in (c). The panels (d), (e), and (f) show the dependence of $\alpha$, $\beta$, and $z$ on $W$, respectively, for the AAM and the RDM in the delocalized phases ($W<1$). The numerical data used for extracting these power exponents are shown in Sec. III of SM~\cite{SM}.
} 
\label{fig2} 
\end{center}
\end{figure*}

{\it Surface-height operator and the roughness.}
To explore dynamical one-parameter scaling, we consider ``quantum surface roughness" defined in Ref.~\cite{height_op2}. 
The essential ingredient is the mathematical analogy between surface growth and one-dimensional nonlinear fluctuating hydrodynamics \cite{Spohn2014,Spohn2016,Mendl_2015,Kulkarni2015}.
The former discusses the dynamics of the surface height $h(x,t)$ that obeys a stochastic partial differential equation, such as the Kardar-Parisi-Zhang (KPZ) equation. For the latter, the recent works \cite{Spohn2014,Spohn2016,Mendl_2015} find that the spatio-temporal correlation function for the sound mode $\phi(x,t)$ shows the dynamical scaling similar to that for $\partial_x h(x,t)$ in the KPZ equation \cite{Spohn2014,Spohn2016,Mendl_2015}.
Similarly, the work \cite{Kulkarni2015} shows that, in the wavenumber and frequency spaces, the correlation function for the local particle number $\rho(x,t)$ in a discrete nonlinear Schrödinger equation well obeys the KPZ scaling. Then, one can see the correspondence between $\partial_x h(x,t)$ and the fluctuation of $\rho(x,t)$. Extending this analogy to quantum systems, we introduce the surface-height operator~\cite{height_op2,height_op3}:

\begin{eqnarray}
\hat{h}_{j} = \sum_{i=1}^{j} \left(  \hat{f}_i^{\dagger} \hat{f}_i  - \nu \right)
\label{sh_op}
\end{eqnarray}
with a filling factor $\nu = N/M$.  The operator represents the particle-number fluctuations summed over the subregion $[1,j]$ and can describe the particle-number fluctuations at a given lengthscale $j$. The averaged surface-height is given by $ h_{\rm av}(t) = \frac{1}{M} \sum_{j=1}^{M}  {\rm Tr}[ \hat{\rho}(t) \hat{h}_j  ]$, where the density matrix $\hat{\rho}(t)$ is averaged over many realizations of the random potentials for the RM and the RDM. 
The surface roughness $w(M,t)$ is defined as the standard deviations of $\hat{h}_j$:
\begin{eqnarray}
w(M,t) = \sqrt{  \frac{1}{M} \sum_{j=1}^{M}  {\rm Tr}[  \hat{\rho}(t)   (\hat{h}_j  -  h_{\rm av}(t)) ^2  ]  }.
\label{main:w}
\end{eqnarray}
As discussed later, the surface roughness is well approximated by the particle-number fluctuations in the half of the system. This implies that the surface roughness measures the correlation between systems divided by two.  

Our previous work~\cite{height_op2} has found that the surface roughness in isolated quantum systems free from disorder exhibits the following FV scaling:
\begin{eqnarray}
w(M,t) = s^{-\alpha} w(sM,s^{z}t) &\propto& 
 \begin{cases}
    t^{\beta} & ( t \ll t_{\rm sat}); \\
    M^{\alpha} & ( t_{\rm sat} \ll t)
  \end{cases}
  \label{FV1}
\end{eqnarray}
with a parameter $s$ and a saturation time $t_{\rm sat}$. Taking $s=1/M$, we obtain $w(M,t) = M^{\alpha} f(t/M^z)$ with a scaling function $f(x)= w(1,x)$. This means that the surface roughness with different $M$ collapses to a single curve after normalization of the ordinate and the abscissa by $M^{\alpha}$ and $M^z$. This dynamical one-parameter scaling is originally discussed in classical systems, and the exponents $\alpha$, $\beta$, and $z$ classify universality of the surface-roughness dynamics \cite{barabasi1995fractal}. The dynamical exponent $z$ satisfies the scaling relation $z=\alpha/\beta$, and $z=1$, $3/2$, and $2$ indicate ballistic, superdiffusive, and diffusive transport. The famous classes are the Edwards-Wilkinson class \cite{OEW} and the KPZ class \cite{OKPZ}, for which the scaling exponents are $(\alpha,\beta,z)=(1/2,1/4,2)$ and $(1/2,1/3,3/2)$, respectively. Our previous work~\cite{height_op2} finds that free fermions (hard-core bosons) without disorders have $(\alpha,\beta,z) \simeq (1/2,1/2,1)$.

{\it Surface-roughness dynamics.}
We numerically investigate the surface roughness to explore the FV scaling in the disordered models. Our numerical method is based on Gaussian states \cite{Cao2019} (see also Sec. II of SM~\cite{SM}). The initial state is a staggered state $\ket{\psi(0)} = \prod_{j=1}^{N} \hat{f}_{2j}^{\dagger} \ket{0}$ with the total particle number $N=M/2$. This initial state has small surface roughness, and thus is suitable to investigate the universal aspect of the surface-roughness growth. 

Figures~\ref{fig2}(a)-(c) show the time evolution of the surface roughness. 
In the delocalized phase ($W=0.5$) of the RDM and the AAM, the surface roughness increases in time and exhibits the FV scaling \eqref{FV1} as shown in Figs.~\ref{fig2}(a) and (b), respectively. The estimated power exponents $(\alpha, \beta, z)$ in the RDM and the AAM are $(0.352, 0.334, 1.01)$ and $(0.487, 0.458,1.02)$, respectively \cite{exponent_evaluation}. These results clearly demonstrate that the dynamical one-parameter scaling indeed exists even in the disordered fermion models. Notably, the exponents in the RDM are anomalous in that they are absent in classical systems and clean systems. This fact is attributed to the LESs in the delocalized phase, as discussed later. 
On the other hand, in the localized phase, the surface roughness is independent of the system size $M$ and does not exhibit clear power-law growth as shown in Figs.~\ref{fig2}(c) for all the models, indicating the absence of the FV scaling. 

We systematically investigate disorder dependence of the exponents $(\alpha, \beta, z)$ by changing $W$ in the delocalized phases. As shown in Figs.~\ref{fig2}(d)-(f), we find that the exponents in the RDM and the AAM are almost independent of $W$. Thus, we conclude that the RDM and the AAM in the delocalized phase show the FV scaling with the exponents $(\alpha, \beta, z) \simeq (0.352, 0.337,1.00)$ and $(0.492, 0.457,1.02)$, respectively, which are obtained by averaging the exponents in Figs.~\ref{fig2}(d)-(f) over $W$. Also, we numerically investigate the dynamics starting from other initial states and find that the choice of the initial state is not important as long as the initial states have small roughness (see Sec. IV of SM~\cite{SM}).

Note that we show the numerical results only for $W \geq 0.3$. 
This is due to the larger localization length for smaller $W$, which makes it difficult to eliminate the finite-size effect.
While we do not have conclusive results for the exponents for small $W$, we conjecture that the exponents are universal for $0<W<1$ in accordance with the phase diagram in Fig.~\ref{fig1}.

The exponents in the AAM are close to $(\alpha, \beta, z) \simeq (0.500, 0.489,1.00)$ for the non-interacting fermion model without disorder~\cite{height_op2}. This coincidence can be understood by considering the numbers of the DLESs and the LESs for the single-particle eigenstate of $\hat{H}$. According to Sec. I of SM \cite{SM}, the numbers of the DLESs and the LESs in the AAM with $W<1$ are proportional to $M$ and $\mathcal{O}(M^0)$, respectively. Thus, we conjecture that the effect of the remaining LESs is too weak and that the exponents are almost the same as the ones for fermion systems without disorder.

\begin{figure}[t]
\begin{center}
\includegraphics[keepaspectratio, width=8.6cm,clip]{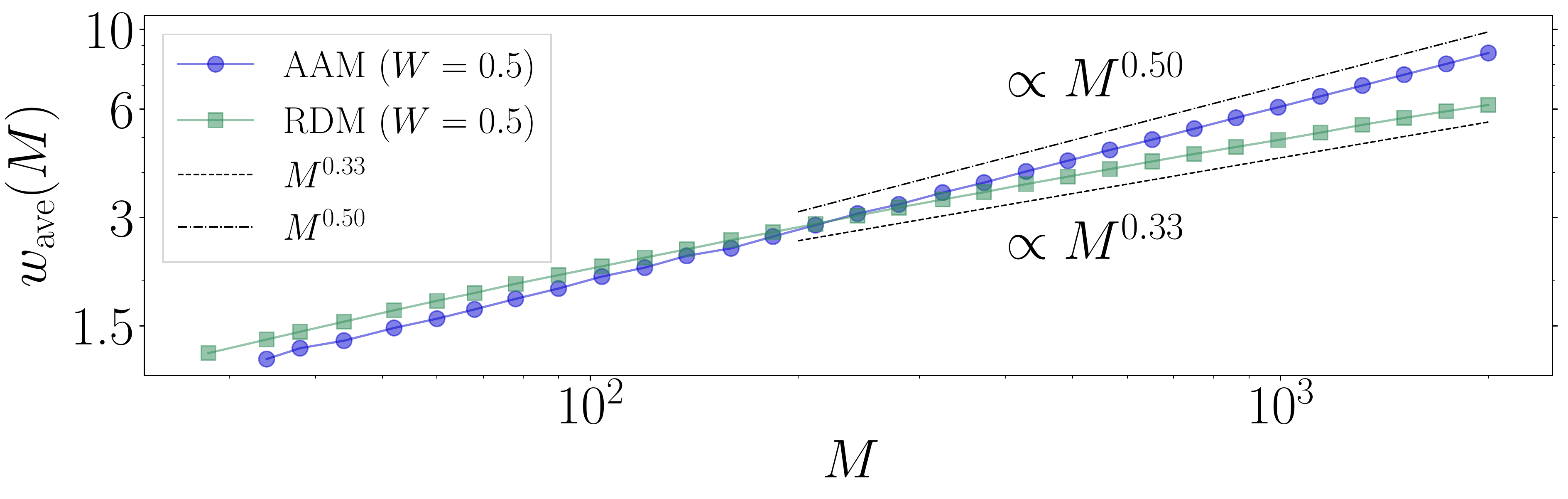}
\caption{
Saturated surface roughness $w_{\rm ave}(M)$ obtained by the approximated diagonal ensemble for the AAM and the RDM with $W=0.5$. 
} 
\label{fig3} 
\end{center}
\end{figure}

The situation drastically changes in the RDM with the anomalous exponent $\alpha \simeq 0.352$. 
According to Ref.~\cite{RDM1} (see also Sec. I of SM \cite{SM}), the numbers of the DLESs and the LESs in the RDM with $W<1$ are proportional to $\sqrt{M}$ and $M$, respectively. In stark contrast to the AAM, the RDM supports many LESs even in the delocalized phase, and they can strongly affect the surface-roughness dynamics. Indeed, just from the information about the eigenstates and the initial state, we can numerically reproduce the exponent $\alpha \simeq 0.33$ and $0.5$ for the RDM and the AAM, respectively, as shown in Fig.~\ref{fig3}. In this calculation, we evaluate the saturated surface roughness $w_{\rm ave}(M)$ using the approximated diagonal ensemble \cite{diagonal1,diagonal2,diagonal3,diagonal4} (see Sec.~V of SM~\cite{SM}). Since we use the same initial states for the RDM and the AAM, the result in Fig.~\ref{fig3} implies that the difference in $\alpha$ originates from the statistical property of the eigenstates. Furthermore, we can analytically derive the non-anomalous exponent $\alpha=0.5$ for systems without LESs, i.e., disorder-free non-interacting systems (see Sec. V of SM~\cite{SM}) and systems satisfying the eigenstate thermalization hypothesis \cite{diagonal1}. All our findings support our argument that the anomalous scaling in the RDM is attributed to the limited number of the DLESs and a large number of LESs.

{\it Entanglement entropy and surface roughness.}
We find that the surface roughness is related to von Neumann EE through a nontrivial relation. 
The EE quantifies quantum entanglement in a pure state in a system divided into two subsystems. Here, we divide the $M$-site system into subsystems $A= \{ j|1\le j\le M/2 \}$ and $B= \{ j|M/2<j \le M \}$ and define the reduced density matrix $\hat{\rho}_{\rm re}(t) = {\rm tr}_{B} [\hat{\rho}_{\rm pure}(t)]$, where $\hat{\rho}_{\rm pure}(t)$ is a density matrix for a single realization of the disordered models. Then, the EE is calculated by $S_{\rm EE}(M,t) = - \overline{ {\rm Tr}_{A} \left[  \hat{\rho}_{\rm re}(t) \log \hat{\rho}_{\rm re}(t) \right] }$, where the overline denotes the ensemble average in the RDM.   

To derive the relation between $S_{\rm EE}(M,t)$ and $w(M,t)$, we assume (i) $h_{\rm av}(t) \simeq 0$, (ii) $w(M,t)^2 \simeq  {\rm Tr} \left[ \hat{\rho}(t)  (\hat{h}_{M/2} - h_{\rm av}(t))^2  \right]$, and (iii) $\sum_{j=1}^{M/2} {\rm Tr} \left[ \hat{\rho}(t)  \hat{n}_{j}  \right] \simeq \nu M/2$. 
The validity of these assumptions is numerically confirmed in Sec. VI of SM \cite{SM}. The assumptions (i) and (ii) lead to 
\begin{eqnarray}
w(M,t)^2  \simeq  {\rm Tr} \left[ \hat{\rho}(t)  \left(  \sum_{j=1}^{M/2} \hat{f}_j^{\dagger} \hat{f}_j  - \frac{M \nu}{2}   \right) ^2  \right]. 
\label{main:bi}
\end{eqnarray} 
Equation~\eqref{main:bi} means that $w(M,t)^2$ can be approximated by the particle-number fluctuation in the half of the system from the averaged number $\nu  M /2$. Thus, both $w(M,t)^2$ and $S_{\rm EE}(M,t)$ have information about the correlation between the divided systems $A$ and $B$. We then find the following relation (see Sec. VI of SM~\cite{SM}):
\begin{eqnarray}
S_{\rm EE}(M,t) &\simeq& 3 w(M,t)^2, 
\label{EE1}
\end{eqnarray}
where we use (iii) and the additional assumption that eigenvalues of the correlation matrix ${\rm Tr}[\hat{\rho}_{\rm pure}(t) \hat{f}_{i}^{\dagger} \hat{f}_j]~(i,j \in A)$ are uniformly distributed between zero and unity. Note that Refs.~\cite{SEE1,SEE2} discuss relations similar to Eq.~\eqref{EE1} for ground states of free-fermion models, but not for dynamics.

\begin{figure}[t]
\begin{center}
\includegraphics[keepaspectratio, width=8.6cm,clip]{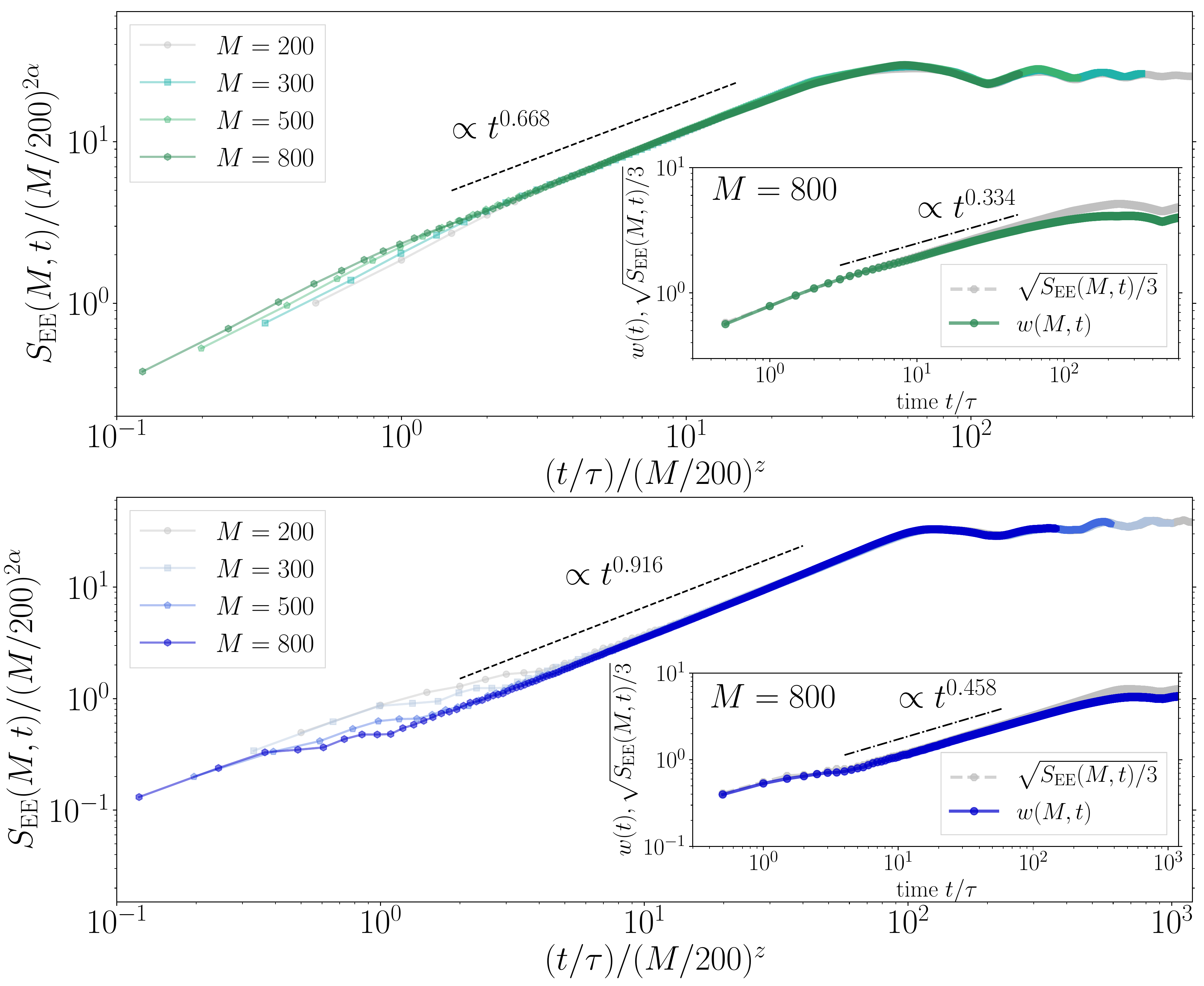}
\caption{FV scaling for von Neumann EE. The upper and lower main panels show $S_{\rm EE}(M,t)$ for the RDM and the AAM, respectively, with $W=0.5$ and $M=200$, $300$, $500$, and $800$, where the time is normalized by $\tau=\hbar/J$, and the ordinate and the abscissa are normalized by $(M/200)^{2\alpha}$ and $(M/200)^z$ with the exponents $\alpha$ and $z$ obtained in Fig.~\ref{fig2}. The insets show time evolutions of $\sqrt{S_{\rm EE}(M,t)/3}$ and $w(M,t)$ for $M=800$, which confirm the success of Eq.~\eqref{EE1} in the early stages of the dynamics.  
} 
\label{fig4} 
\end{center}
\end{figure}

Substituting Eq.~\eqref{EE1} into Eq.~\eqref{FV1}, we obtain the FV-type scaling in the delocalized phases:
\begin{eqnarray}
S_{\rm EE}(M,t) = s^{-2\alpha} S_{\rm EE}(sM,s^{z}t) &\propto& 
 \begin{cases}
    t^{2\beta} & ( t \ll t_{\rm sat}); \\
    M^{2\alpha} & ( t_{\rm sat} \ll t).
  \end{cases}
  \label{EE_FV}
\end{eqnarray}
Figure~\ref{fig4} shows time evolutions of $S_{\rm EE}(M,t)$ in the RDM and the AAM with $W=0.5$. Our numerical results clearly reveal that the EE well obeys the FV-type scaling \eqref{EE_FV}. The insets of Fig.~\ref{fig4} compare both of the sides of Eq.~\eqref{EE1}, showing that the relation works quite well especially in the early stages of the dynamics. Although they deviate from each other in the late stages, the FV-type scaling still holds with the expected exponent $(2\alpha,2\beta,z)$. 

This finding suggests that the surface roughness may become a possible measure for entanglement and its universal scaling. Furthermore, we rigorously prove in Sec. VI of SM~\cite{SM} that if the bipartite number fluctuation ${\rm Tr}[\hat{\rho}(t) \hat{h}_{M/2}^2]$ \cite{bipartite1,bipartite2,bipartite3} with the assumption (iii) exhibits power-law growth $ t^{\beta}$, $S_{\rm EE}(M,t)$ also grows as $t^{2\beta}$ in the thermodynamic limit (and vice versa). 

Finally, we comment on the entanglement dynamics studied in view of the surface roughness. 
Using quantum circuit models, Nahum $et~al.$ \cite{Nahum2,Nahum3} show that the EE obeys the KPZ equation. This means that the EE itself behaves as surface height, which are different from our result of Eq.~\eqref{EE1}. 
The difference may be attributed to the models used in the previous and our works because they have different conserved quantities, which can lead to the distinct long-time dynamics. 
We also stress that the FV scaling has not been observed in Refs.~\cite{Nahum2,Nahum3} in that they do not examine saturation of the fluctuations of the EE.

{\it Conclusion and outlook.}
We have numerically found the dynamical one-parameter scaling of surface roughness and entanglement entropy in the disordered fermion models, including anomalous scaling arising from the partial quantum localization \cite{comment1}. 
Our study opens an unexplored avenue for pursuing unexpected relation between Anderson localization and surface growth physics 
through the FV scaling and the EE. From this viewpoint, it is interesting to investigate universality class of the FV scaling in many-body localization \cite{MBL0,MBL1,MBL2,MBL3,MBL4,MBL5,MBL6,MBL7,MBL8,MBL9,MBL10,MBL11,MBL12,MBL13} and the Anderson localization with long-range interactions.

\begin{acknowledgments}
We would like to thank K. Kawabata, X. Chai, D. Lao, and C. Raman for fruitful discussions.
This work was supported by JST-CREST (Grant No. JPMJCR16F2), JSPS KAKENHI (Grant Nos. JP18K03538, JP19H01824,  JP19K14628, and 20H01843), Foundation of Kinoshita Memorial Enterprise, and the Program for Fostering Researchers for the Next Generation (IAR, Nagoya University) and Building of Consortia for the Development of Human Resources in Science and Technology (MEXT). 
\end{acknowledgments}

\bibliography{reference}

\begin{thebibliography}{85}
\bibitem [{exp()}]{exponent_evaluation}%
  \BibitemOpen
  \href@noop {} {\bibinfo  {journal} {While exponents ($\alpha$, $\beta$) and
  $z$ are independently determined, one can still see that the relation
  $z=\alpha/\beta$ is well satisfied}\ }\BibitemShut {NoStop}%
\end{thebibliography}

\widetext
\clearpage

\setcounter{equation}{0}
\setcounter{figure}{0}
\setcounter{section}{0}
\setcounter{table}{0}
\renewcommand{\theequation}{S-\arabic{equation}}
\renewcommand{\thefigure}{S-\arabic{figure}}
\renewcommand{\thetable}{S-\arabic{table}}

\section*{Supplemental Material for ``Anomalous Dynamical Scaling of Roughness in Disordered Fermion models''}
This supplemental material describes the following topics:
\begin{itemize}
\item[  ]{ (I) Numbers of delocalized and localized eigenstates in the disordered fermion models, }
\item[  ]{ (II) Numerical method, } 
\item[  ]{ (III) Numerical data for the surface-roughness dynamics, }
\item[  ]{ (IV) Dependence of the surface roughness dynamics on initial states, }
\item[  ]{ (V) Family-Vicsek-scaling exponent $\alpha$ in the approximated diagonal ensemble,  }
\item[  ]{ (VI) Relation between the von Neumann entanglement entropy and the surface roughness,}
\item[  ]{ (VII) Anomalous behavior of single-particle transport in the RDM,}
\item[  ]{ (VIII) Experimental possibility. }
\end{itemize}

\section{Numbers of delocalized and localized eigenstates in the disordered fermion models \label{DLS}}
This section addresses numbers of delocalized eigenstates (DLESs) and localized eigenstates (LESs) in the three disordered fermion models, namely the random model (RM), the random-dimer model (RDM), and the Aubry-André model (AAM). First, we numerically solve the stationary Schrödinger equation in the single-particle Fock basis $\{ \hat{f}_{i}^{\dagger}\ket{0} |  ~i=1,2, \cdots, M  \}$:
\begin{eqnarray}
\sum_{j=1}^{M} H_{ij} v_{j\alpha} = \epsilon_{\alpha} v_{i\alpha}, 
\label{App1}
\end{eqnarray}
where $v_{i \alpha}$ and $\epsilon_{\alpha}$ are an eigenvector and an eigenvalue labeled by a quantum number ${\alpha}=1,2, \cdots, M$. 
The matrix element $H_{ij} $ is given by
\begin{eqnarray}
  H = \left(
    \begin{array}{ccccccc}
       V_1     &  -J        &              &              &             & A    &\\
      -J         &   V_2    &   -J        &              &              &        &  \\
                  &  -J        &   V_3     &   -J        &              &        &  \\
                  &             & \ddots   & \ddots  & \ddots   &        & \\
                  &             &              & \ddots  & \ddots   &  \ddots  & \\
          A      &             &              &             &       -J     &   V_M    & 
    \end{array}
  \right). 
\label{Hmat}
\end{eqnarray}
Here, $A$ is a constant depending on the boundary conditions, and becomes $0$ and $-J$ in the open and periodic boundary condition, respectively. Next, we define the inverse participation ratio $R$ as
\begin{eqnarray}
 R_{\alpha} = \frac{\displaystyle \sum_{j=1}^{M} |v_{j \alpha}|^4}{ \displaystyle \left( \sum_{j=1}^{M} |v_{j \alpha}|^2 \right)^2}.
\label{App2}
\end{eqnarray}
If an eigenvector is delocalized, the ratio $R_{\alpha}$ is proportional to $1/M$. On the other hand, $R_{\alpha}$ becomes $\mathcal{O}(1)$ if an eigenvector is localized. In this work, we identify DLESs by the condition that $R_{\alpha}$ is smaller than $10/M$, and count the number $N_{\rm DLES}(M)$ of the DLESs. Then, the number of the LESs is defined by $N_{\rm LES}(M):= M - N_{\rm DLES}(M)$.

\begin{figure}[t]
\begin{center}
\includegraphics[keepaspectratio, width=18cm,clip]{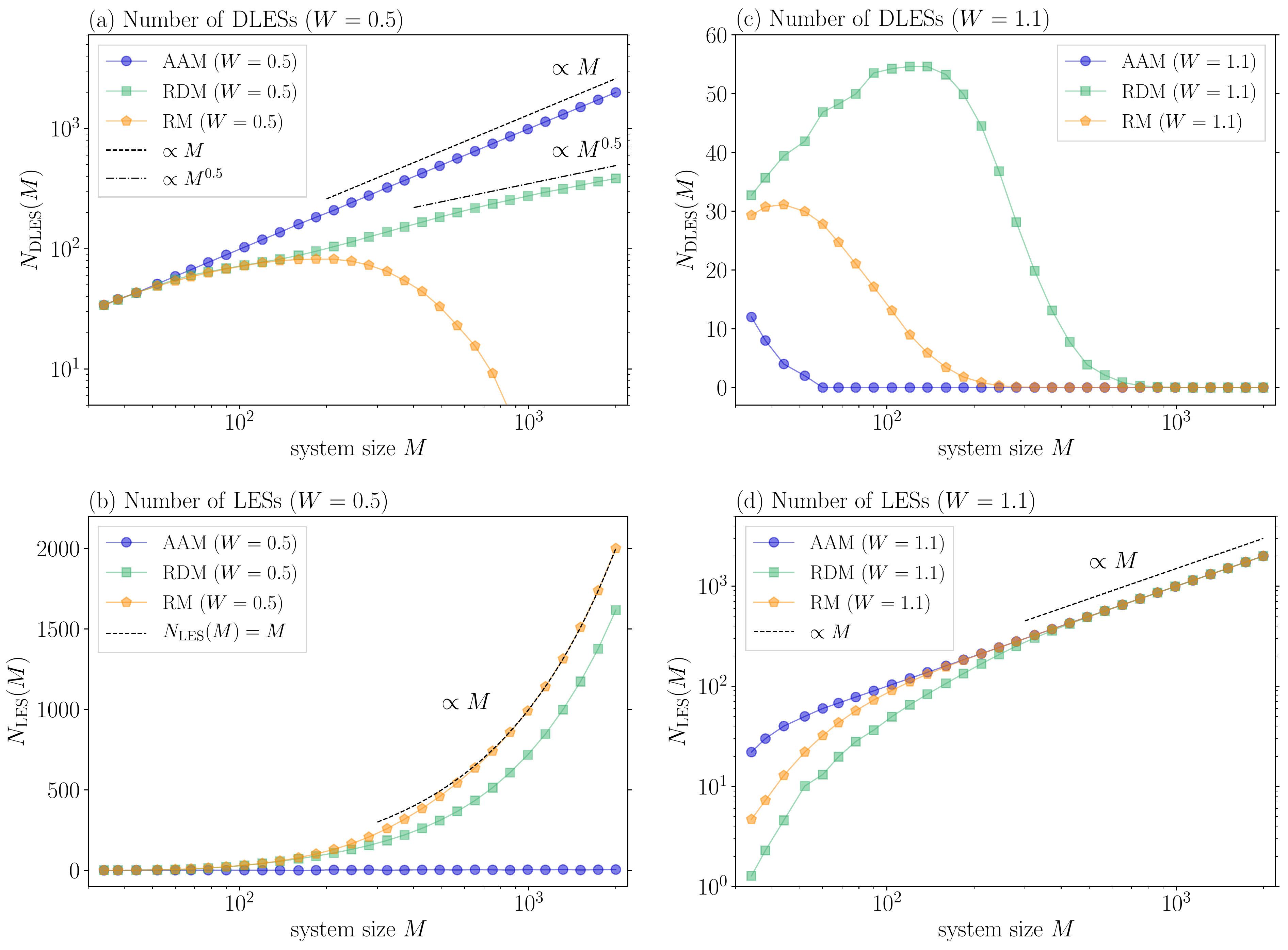}
\caption{(a,b) $N_{\rm DLES}(M)$ and $N_{\rm LES}(M)$ in the RDM, the AAM, and the the RM with $W=0.5$.
The numbers of the DLESs in the RDM and the AAM obey the $\sqrt{M}$ and $M$ power-law growth, while one in the RM does not increase. 
On the one hand, the numbers of the LESs in the RDM and the RM grows with increasing $M$, while one in the AAM is almost to zero. 
(c,d) $N_{\rm DLES}(M)$ and $N_{\rm LES}(M)$ in the RDM, the AAM, and the RM with $W=1.1$. The numbers of the DLESs approach to zero in large $M$, while those of the LESs grows with the power law $N_{\rm LES}(M) \propto M $ in large $M$. 
} 
\label{Num_DLS} 
\end{center}
\end{figure}

Figure~\ref{Num_DLS} shows $N_{\rm DLES}(M)$ and $N_{\rm LES}(M)$ in the RM, the RDM, and the AAM. The left and right panels are the results for $W=0.5$ and $W=1.1$. The RDM with $W=0.5$ is in the delocalized phase, and the numbers of the DLESs and the LESs are proportional to $\sqrt{M}$ and $M$, respectively. This result is consistent with Ref.~\cite{RDM1}. On the other hand, the AAM shows $N_{\rm LES}(M)  \ll 1$, which means $N_{\rm DLES}(M) \simeq M$. Thus, most of the eigenstates are extended to the entire system, and the LESs do not affect the surface-growth dynamics. In the localized phase, $N_{\rm DLES}(M)$ and $N_{\rm LES}(M)$ in all the models approaches to $\mathcal{O}(1)$ and $M$, respectively, when $M$ is much larger than unity.   

\begin{figure}[t]
\begin{center}
\includegraphics[keepaspectratio, width=18cm,clip]{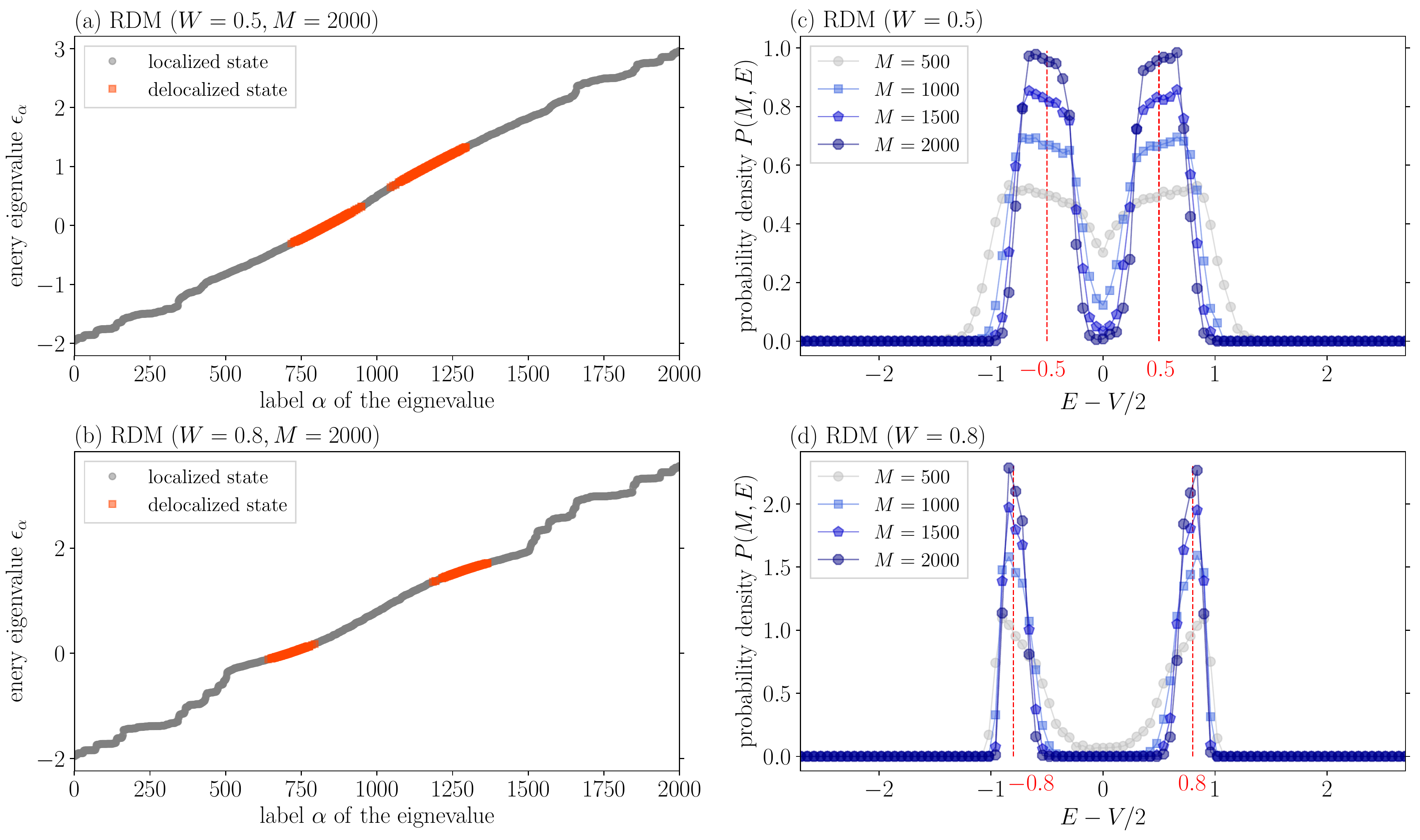}
\caption{
Distributions of the delocalized and localized eigenstates in the RDM with $W=0.5$ and $0.8$. 
Numerically diagonalizing the Hamiltonian of the RDM, we obtain all the eigenstates and classify the delocalized and localized states by calculating the inverse participation ratio. Figures (a) and (b) show the energy eigenvalues for a single realization of the random-dimer potential in the case of $W=0.5$ and $0.8$, respectively. Here, the system size is set to be $M=2000$, and the eigenvalues are labeled in the ascending order. The orange and gray makers show the eigenvalues for the delocalized and localized eigenstates. 
Figures (c) and (d) show the probability density function for the existence of the delocalized states as a function of the energy $E$, where we subtract $V/2$ in the horizontal axis to make the averaged potential energy zero. The system size is $M=500, 1000, 1500, $ and $2000$.
The red lines denote the resonant energies of the RDM. The probability density functions are calculated by the average over many realizations of the random-dimer potentials.
} 
\label{mobility_edge} 
\end{center}
\end{figure}

We numerically investigate where the delocalized eigenstates exist in the RDM with $W<1$. 
Figures~\ref{mobility_edge}(a) and (b) show the distributions of the localized and delocalized eigenstates for $W=0.5$ and $0.8$, respectively.
Here, the eigenvalues are labeled in the ascending order and the distributions are obtained by single realizations of the random-dimer potential. 
We can see that the localized and delocalized states are well separated in energy.
Taking the ensemble average for the random-dimer potential, we plot the probability density functions for the delocalized states in Figs.~\ref{mobility_edge} (c) and (d), which clearly exhibit  peak structures around $E = \frac{V}{2} \pm W$. 
The peak structure can be intuitively understood by considering existence of perfect transmission in the RDM. 
As discussed in Ref.~\cite{RDM1}, the random-dimer potential allows a particle at a specific energy to transport without any reflection, and the energies of the peaks in Figs.~\ref{mobility_edge} (c) and (d) correspond to this resonant energy. 
This resonant behavior shown here was reported in Ref.~\cite{RDM_mobility}: using the transfer matrix, the paper systematically investigated the width of the resonant peaks by changing the system size and found that the width becomes narrow with increasing the system size. In that sense, the authors argued that mobility edges do not exist in the RDM in the thermodynamic limit. 
As shown in Figs.~\ref{mobility_edge} (c) and (d), our numerical results find the similar behavior that the width of the peaks decreases as the system size $M$ increases.

\clearpage

\section{Numerical method}
To solve the Schrödinger equation in non-interacting fermion models, we use the numerical method described in Ref.~\cite{Cao2019}.
In this section, we review the details of how to implement it.  

The system considered here is $N$-fermions on a one-dimensional lattice. 
We denote by $\hat{f}_j$ and $\hat{f}^{\dagger}_j~(j=1,\cdots,M)$ the fermionic annihilation and creation operators at a site $j$ with the number $M$ of the lattice sites, respectively. 
Then, all the Hamiltonians used in this work are given by the following quadratic form:
\begin{eqnarray}
\hat{H} = \sum_{i,j=1}^{M} \hat{f}_{i}^{\dagger} H_{ij} \hat{f}_j
\label{sup:H}
\end{eqnarray}
with an $M\times M$ hermitian matrix $H_{ij}$. 
Numerically solving the Schrödinger equation with Eq.~\eqref{sup:H}, we utilize the Gaussian state defined by
\begin{eqnarray}
\prod_{\alpha=1}^{N} \left( \sum_{j=1}^{M} \hat{f}^{\dagger}_j G_{j \alpha} \right) \ket{0},  
\label{sup:G}
\end{eqnarray}
where $G_{j \alpha}$ is a $M \times N$ matrix and $\ket{0}$ is a vacuum state satisfying $\hat{f}_{j} \ket{0} = 0$ for $\forall j$. 
Here, the matrix is assumed to satisfy $ \sum_{j} G_{j \alpha}^* G_{j \beta} = \delta_{\alpha \beta}$, which ensures that the operator $\sum_{j=1}^{M} \hat{f}^{\dagger}_j G_{j \alpha}$ obeys the fermionic anticommutator relation.  
The unitary time evolution with Eq.~\eqref{sup:H} can keep this Gaussian-state structure in time if an initial state is the Gaussian state. 
Thus, we can track the time-evolved state only by calculating the matrix $G_{j \alpha}$.

We here show that the Gaussian-state structure is kept in time under Eq.~\eqref{sup:H}. 
The initial state $\ket{\psi(0)}$ is assumed to be the Gaussian state:
\begin{eqnarray}
\ket{\psi(0)} = \prod_{\alpha=1}^{N} \left( \sum_{j=1}^{M} \hat{f}^{\dagger}_j U_{j \alpha} (0) \right) \ket{0},  
\label{initial_state_s1}
\end{eqnarray}
where $U_{j \alpha} (0)$ is a $M \times N$ matrix. 
In the main text, we use $U_{j\alpha}(0)=\delta_{j,2\alpha}$ corresponding to a staggered state $\prod_{j=1}^{M/2} \hat{f}^{\dagger}_{2j} \ket{0}$.
Applying the unitary operator $\hat{U}(t) = \exp{(-i\hat{H}t/ \hbar)}$ to Eq.~\eqref{initial_state_s1}, we obtain the time-evolved state $\ket{\psi(t)}$ at a time $t$: 
\begin{eqnarray}
\ket{\psi(t)} &=& \hat{U}(t) \prod_{\alpha=1}^{N} \left( \sum_{j=1}^{M} \hat{f}^{\dagger}_j U_{j \alpha} (0) \right) \ket{0} \\
&=&  \prod_{\alpha=1}^{N} \left( \sum_{j=1}^{M} \hat{U}(t) \hat{f}^{\dagger}_j \hat{U}^{\dagger}(t) U_{j \alpha}(0) \right) \ket{0}. 
\label{state_s1}
\end{eqnarray}
Here, we use $\hat{U}(t) \hat{U}^{\dagger}(t)=1$ and $\hat{H} \ket{0} = 0$ to derive the last line.
Calculating $\hat{U}(t) \hat{f}^{\dagger}_j \hat{U}^{\dagger}(t)$ by means of the Baker-Campbell-Hausdorff formula, we find
\begin{eqnarray}
\hat{U}(t) \hat{f}^{\dagger}_j \hat{U}^{\dagger}(t) = \sum_{k=1}^{M} \hat{f}_k^{\dagger} A_{kj}(t)
\label{state_s2}
\end{eqnarray}
with coefficients $A_{kj}(t)~(j,k=1,\cdots,M)$ depending on $H_{ij}$. As a result, the quantum state~\eqref{state_s1} is expressed by
\begin{eqnarray}
\ket{\psi(t)} = \prod_{\alpha=1}^{N} \left( \sum_{k=1}^{M} \hat{f}^{\dagger}_k U_{k \alpha}(t) \right) \ket{0}
\label{state_s3}
\end{eqnarray}
with $U_{k \alpha}(t) = \sum_{j=1}^{M} A_{kj}(t) U_{j \alpha} (0)$. Obviously, Eq.~\eqref{state_s3} has the Gaussian-state structure~\eqref{sup:G}. 
Thus, we can investigate the quantum dynamics only by calculating the time evolution of $U_{k \alpha}(t)$ starting from the initial coefficient $U_{k \alpha}(0)$. In all our numerical simulations, we calculate $U_{k \alpha}(t)$ using the Crank-Nicolson method, which conserves the norm $\braket{\psi(t) | \psi(t)}=1$ at least up to the order of $10^{-7}$. 

Finally, we comment on a correlation matrix. 
Using Eq.~\eqref{state_s3}, we express the correlation matrix $D_{ij}(t)$ with $U_{i \alpha}(t)$:
\begin{eqnarray}
D_{ij}(t) &:=& {\rm Tr} \left[ \hat{\rho}(t) \hat{f}_i^{\dagger} \hat{f}_j \right] \\
&=& \sum_{\alpha=1}^{N} U_{i \alpha}^*(t) U_{j \alpha}(t).
\label{correlation_M1}
\end{eqnarray}
By definition, the occupation number at a site-$j$ is given by $ D_{jj}= \braket{ \hat{f}_j^{\dagger} f_j } $. 
Similarly, we can derive
\begin{eqnarray}
 {\rm Tr} \left[ \hat{\rho}(t) \hat{f}_i^{\dagger} \hat{f}_i \hat{f}_j^{\dagger} \hat{f}_j \right] = 
 \begin{cases}
    D_{ii} D_{jj} - D_{ij} D_{ji} & (i \neq j); \\
    D_{ii} & (i=j).
  \end{cases}
\label{correlation_M2}
\end{eqnarray}
We numerically calculate Eqs.~\eqref{correlation_M1} and \eqref{correlation_M2}, and then obtain the surface roughness investigated in the main text.

\section{Numerical data for the surface-roughness dynamics}
\begin{figure*}[b]
\begin{center}
\includegraphics[keepaspectratio, width=18.0cm,clip]{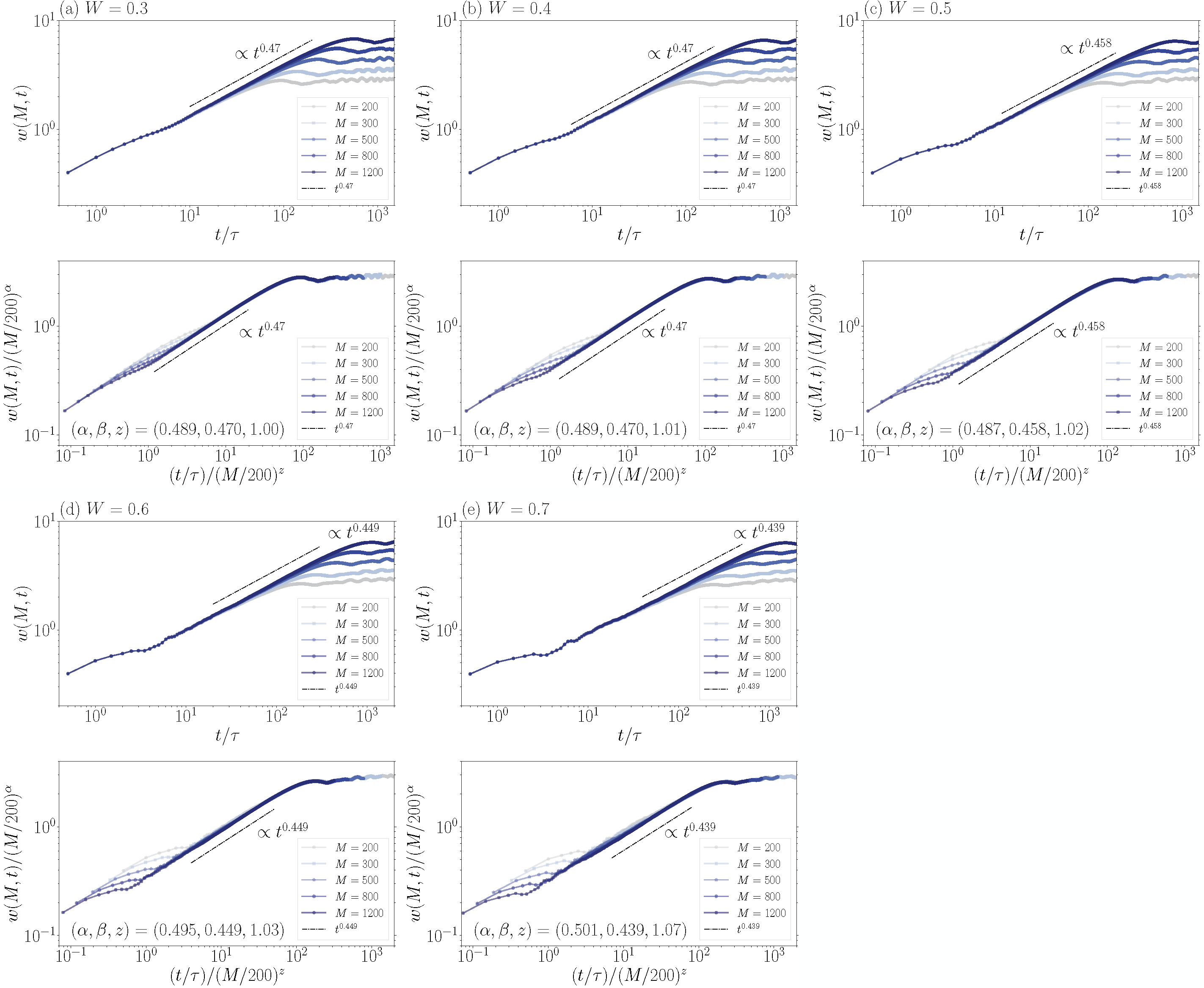}
\caption{ Surface-roughness dynamics in the AAM with (a) $W=0.3$, (b) $W=0.4$, (c) $W=0.5$, (d) $W=0.6$, and (e) $W=0.7$. The upper and lower panels show the raw data and the data with the normalized ordinate and abscissa by $(M/200)^{\alpha}$ and $(M/200)^{z}$, respectively. The extracted exponents are $(\alpha,\beta,z)=(0.480,0.470,1.00), (0.489,0.470,1.01), (0.487,0.458,1.02), (0.495,0.449,1.03)$, and $(0.501,0.439,1.07)$ for (a), (b), (c), (d), and (e), respectively.
} 
\label{S_AAM} 
\end{center}
\end{figure*}

\begin{figure*}[b]
\begin{center}
\includegraphics[keepaspectratio, width=18.0cm,clip]{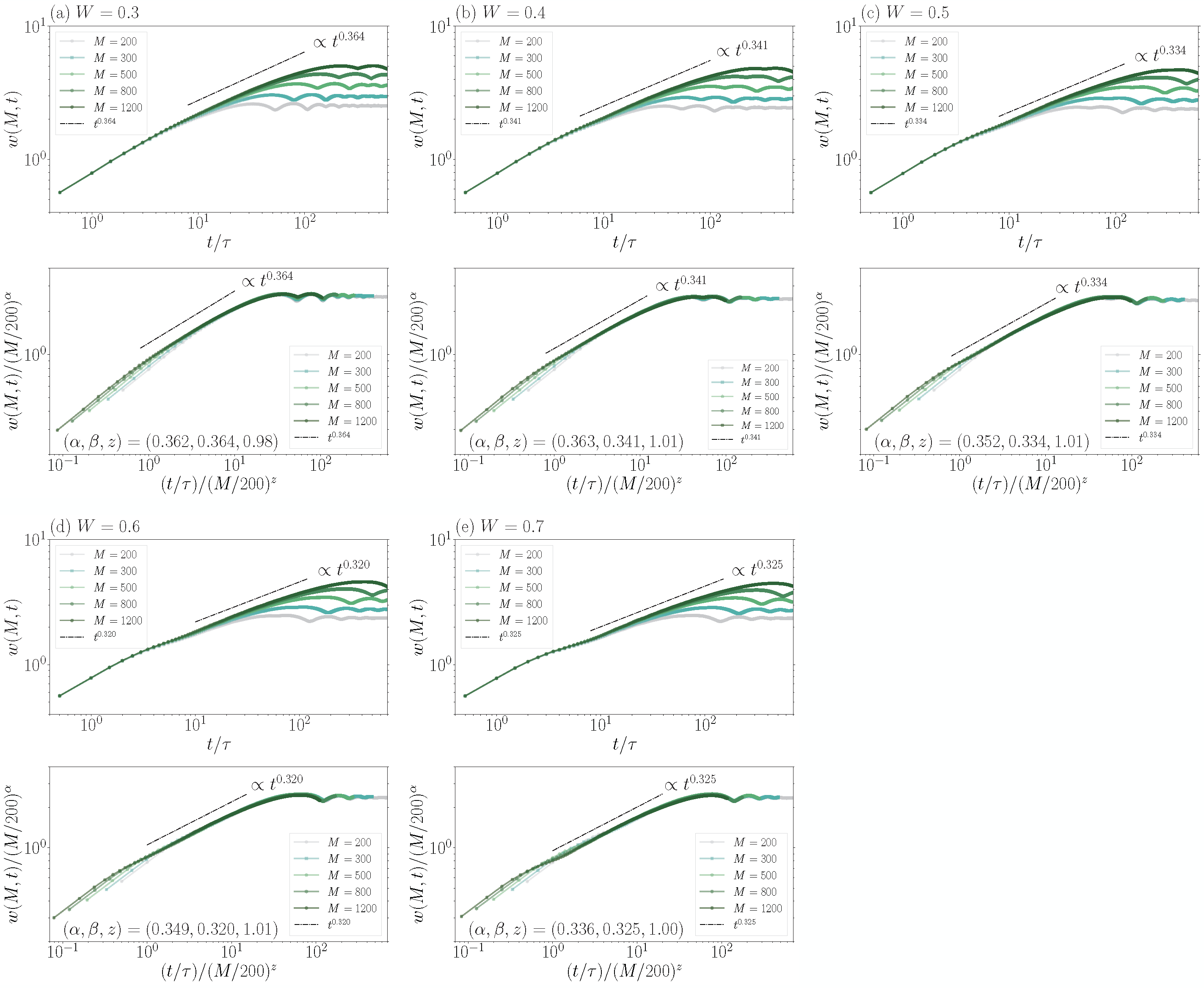}
\caption{ Surface-roughness dynamics in the RDM with (a) $W=0.3$, (b) $W=0.4$, (c) $W=0.5$, (d) $W=0.6$, and (e) $W=0.7$. The upper and lower panels show the raw data and the data with the normalized ordinate and abscissa by $(M/200)^{\alpha}$ and $(M/200)^{z}$, respectively.
The extracted exponents are $(\alpha,\beta,z)=(0.362,0.364,0.98),(0.363,0.341,1.01),(0.352,0.334,1.01),(0.349,0.32,1.01)$, and $(0.336,0.325,1.00)$ for (a), (b), (c), (d), and (e), respectively.
} 
\label{S_RDM} 
\end{center}
\end{figure*}

We show all the numerical results for time evolution of the surface roughness, which are used to extract the universal power exponents featuring the Family-Vicsek (FV) scaling in Figs.~2 and 3 of the main text. Figures~\ref{S_AAM} and \ref{S_RDM} show time evolution of all the surface roughness in the AAM and the RDM with $W<1$, respectively. Our method for calculating the power exponents is based on Ref.~\cite{height_op2,expoents_cal}, and the tables~\ref{tab:1} and \ref{tab:2} summarize the time region used for the calculation.  

\begin{table}[b]
\begin{ruledtabular}
\begin{tabular}{ l c  c  c c}
\textrm{$W$}&
\textrm{system size $M$}& 
\textrm{time region for $\alpha$ and $z$}&
\textrm{time region for $\beta$}&
\\
\colrule
0.3 & $200,300,500,800,1200$ &  $[15\tau, 160\tau]$ & $[15\tau, 100\tau]$ \\
0.4 & $200,300,500,800,1200$ &  $[15\tau, 180\tau]$ & $[15\tau, 100\tau]$  \\
0.5 & $200,300,500,800,1200$ &  $[15\tau, 240\tau]$ & $[15\tau, 150\tau]$  \\
0.6 & $       300,500,800,1200$ &  $[40\tau, 400\tau]$ & $[40\tau, 300\tau]$  \\
0.7 & $              500,800,1200$ &  $[40\tau, 850\tau]$ & $[40\tau, 400\tau]$  \\
\end{tabular}
\end{ruledtabular}
\caption{\label{tab:1}%
Fitting information for the AAM in the delocalized phase ($W<1$). 
The first column denotes the disorder strength $W$, and second and third ones show the system sizes $M$ and the fitting time regions used for the evaluation of $\alpha$ and $z$. To extract the exponent $\beta$, we use the data for $M=1200$ and the time region given in the fourth column.
}
\end{table}

\clearpage


\begin{table}[b]
\begin{ruledtabular}
\begin{tabular}{ l c  c  c c}
\textrm{$W$}&
\textrm{system size $M$}& 
\textrm{time region for $\alpha$ and $z$}&
\textrm{time region for $\beta$}&
\\
\colrule
0.3 & $500,800,1200$ &  $[7\tau, 130\tau]$ & $[7\tau, 35\tau]$ \\
0.4 & $300,500,800,1200$ &  $[7\tau, 80\tau]$ & $[7\tau, 50\tau]$  \\
0.5 & $300,500,800,1200$ &  $[8\tau, 120\tau]$ & $[8\tau, 50\tau]$  \\
0.6 & $200,300,500,800,1200$ &  $[10\tau, 130\tau]$ & $[10\tau, 60\tau]$  \\
0.7 & $200,300,500,800,1200$ &  $[12\tau, 140\tau]$ & $[12\tau, 60\tau]$  \\
\end{tabular}
\end{ruledtabular}
\caption{\label{tab:2}%
Fitting information for the RDM in the delocalized phase ($W<1$). 
The first column denotes the disorder strength $W$, and second and third ones show the system sizes $M$ and the fitting time regions used for the evaluation of $\alpha$ and $z$. To extract the exponent $\beta$, we use the data for $M=1200$ and the time region given in the fourth column. }
\end{table}

\clearpage

\section{Dependence of the surface roughness dynamics on initial states}
We numerically investigate the surface roughness dynamics starting from three initial states different from the staggered state used in the main text. 
Here, the system size and the disorder strength are set to be $M=800$ and $W=0.5$. 
The initial states considered here are given by 
\begin{eqnarray}
\ket{\psi (0)} = \prod_{j=0}^{(M-8)/8} \hat{P}_{8j + 1} \ket{0}, 
\end{eqnarray}
where an operator $\hat{P}_{8j+1}~(j=0, \cdots, (M-8)/8)$ take the following form depending on the four initial states:
\begin{eqnarray}
\hat{P}_k = 
\displaystyle
 \begin{cases}
 \displaystyle
    \hat{f}^{\dagger}_{k+1} \hat{f}^{\dagger}_{k+3} \hat{f}^{\dagger}_{k+5} \hat{f}^{\dagger}_{k+7} & ({\rm initial~state~1}); \\
    \\
    \hat{f}^{\dagger}_{k} \hat{f}^{\dagger}_{k+1} \hat{f}^{\dagger}_{k+4} \hat{f}^{\dagger}_{k+5} & ({\rm initial~state~2}); \\
    \\
    \hat{f}^{\dagger}_{k} \hat{f}^{\dagger}_{k+3} \hat{f}^{\dagger}_{k+4} \hat{f}^{\dagger}_{k+6} & ({\rm initial~state~3}); \\
    \\
    \hat{f}^{\dagger}_{k} \hat{f}^{\dagger}_{k+1} \hat{f}^{\dagger}_{k+2} \hat{f}^{\dagger}_{k+6} & ({\rm initial~state~4}).
  \end{cases}
\label{S_unit}
\end{eqnarray}
The initial states 2, 3,  and 4 are different from the staggered state (initial state 1). 
They all have small surface roughness and are suitable for investigating the growth dynamics.
 
\begin{figure*}[b]
\begin{center}
\includegraphics[keepaspectratio, width=18.0cm,clip]{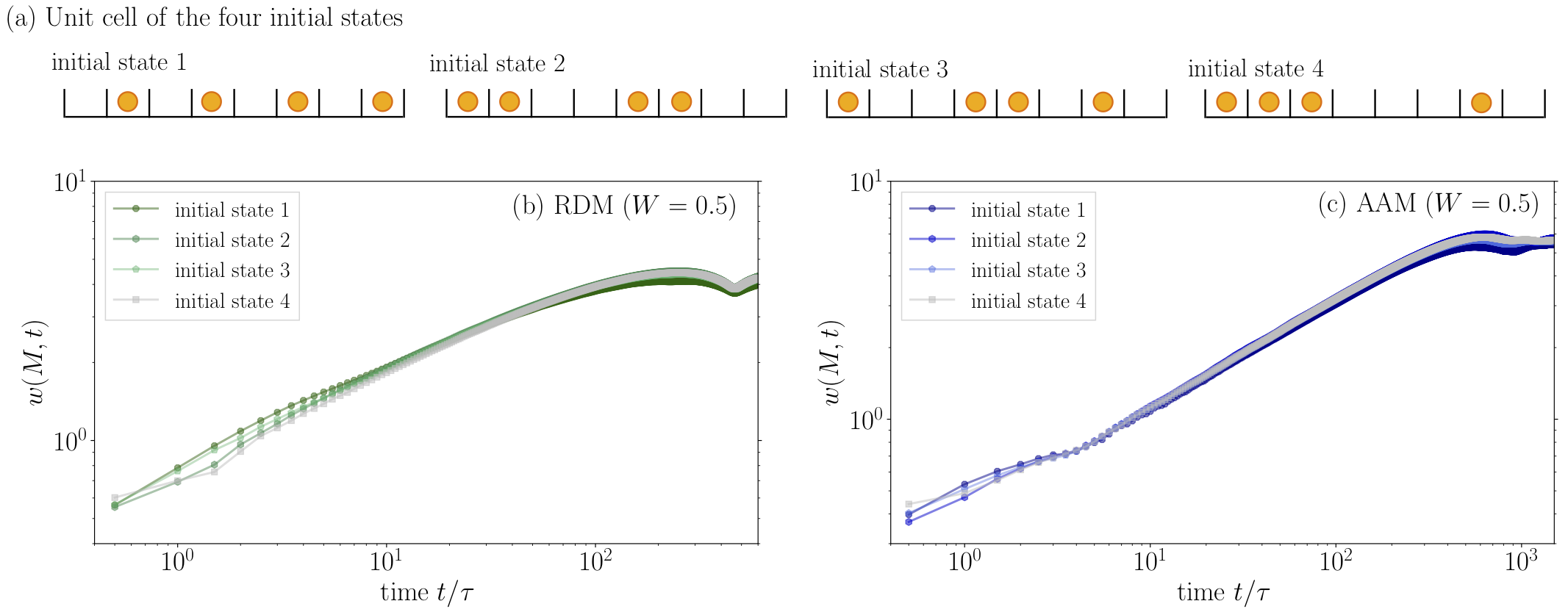}
\caption{
Dependence of the surface roughness dynamics on the four initial states.
(a) Fock-state configurations for the unit cells of four different initial states. The unit cells (8 lattice points) of the initial states 1, 2, 3, and 4 have four particles denoted by yellow circles. (b, c) Time evolution of the surface roughness for the RDM and the AAM in the delocalized phases ($W=0.5$). The figures show that the time evolution is almost independent of the initial states especially in $t/\tau > 10$. 
} 
\label{S_initial_state} 
\end{center}
\end{figure*}

Figure~\ref{S_initial_state}~(a) shows the four configurations of the Fock states corresponding to Eq.~\eqref{S_unit}. 
Using these initial states, we numerically calculate the time evolution of the surface roughness in the RDM and the AAM as shown in Figs.~\ref{S_initial_state}~(b) and (c). We find that the time evolution is almost independent of the choice of the initial states for $t/\tau > 10$, for which the power-law growth emerges. Thus, we argue that the FV scaling exponent is universal as long as the initial states have small surface roughness.  

\clearpage

\section{Family-Vicsek-scaling exponent $\alpha$ in the approximated diagonal ensemble}
Employing an approximated diagonal ensemble \cite{diagonal1,diagonal2,diagonal3,diagonal4}, we numerically obtain $\alpha = 0.33$ and $0.50$ in the RDM and the AAM without directly solving the time-dependent Schrödinger equation.
This section explains the detailed calculations. 

\subsection{Diagonalization of the Hamiltonian}
In this subsection, we diagonalize the quadratic Hamiltonian~\eqref{Hmat} and give a relation between the bare-fermion operators $\{ \hat{f}_i,\hat{f}_i^{\dagger} \}_{i=1,\cdots,M}$ and the quasi-fermion ones $\{ \hat{F}_{\alpha},\hat{F}_{\alpha}^{\dagger} \}_{\alpha=1,\cdots,M}$ (see Eqs.~\eqref{quasi1} and \eqref{quasi2}).
Solving an eigenvalue problem with the Hermitian matrix $H$ of Eq.~\eqref{Hmat}, we obtain the eigenvalue $\epsilon_{\alpha}$ and the corresponding eigenvector $\bm{v}_{\alpha}$ with the label $\alpha=1,2, \cdots, M$. Then, the Hamiltonian is diagonalized as 
\begin{eqnarray}
V^{\dagger} H V = {\rm diag}(\epsilon_1, \epsilon_2, \cdots, \epsilon_M).
\end{eqnarray}
Here, we define a unitary matrix $V = (\bm{v}_1, \bm{v}_2, \cdots, \bm{v}_M)$.
Using all these results, we finally obtain
\begin{eqnarray}
\hat{H} = \sum_{\alpha =1}^{M}  \epsilon_{\alpha} F^{\dagger}_{\alpha}  \hat{F}_{\alpha}, 
\end{eqnarray}
where $\hat{F}_{\alpha}$ and $\hat{F}_{\alpha}^{\dagger}$ are annihilation and creation operators for the quasi-fermions defined by
\begin{eqnarray}
\hat{F}_{\alpha} = \sum_{j=1}^{M} v^*_{j\alpha} \hat{f}_{j},  \label{quasi1} \\
\hat{F}_{\alpha}^{\dagger} = \sum_{j=1}^{M} v_{j\alpha} \hat{f}_{j} ^{\dagger}. \label{quasi2}
\end{eqnarray}
The inverse transformations for Eqs.~\eqref{quasi1} and \eqref{quasi2} become
\begin{eqnarray}
\hat{f}_{j} = \sum_{\alpha=1}^{M} v_{j\alpha} \hat{F}_{\alpha}, \\
\hat{f}_{j}^{\dagger} = \sum_{\alpha=1}^{M} v_{j\alpha}^* \hat{F}_{\alpha} ^{\dagger}.
\end{eqnarray}
In what follows, employing the transformations, we use the diagonal ensemble to investigate the surface roughness in the stationary state. 

\subsection{Approximated expression of the surface roughness}
\begin{figure}[t]
\begin{center}
\includegraphics[keepaspectratio, width=18cm,clip]{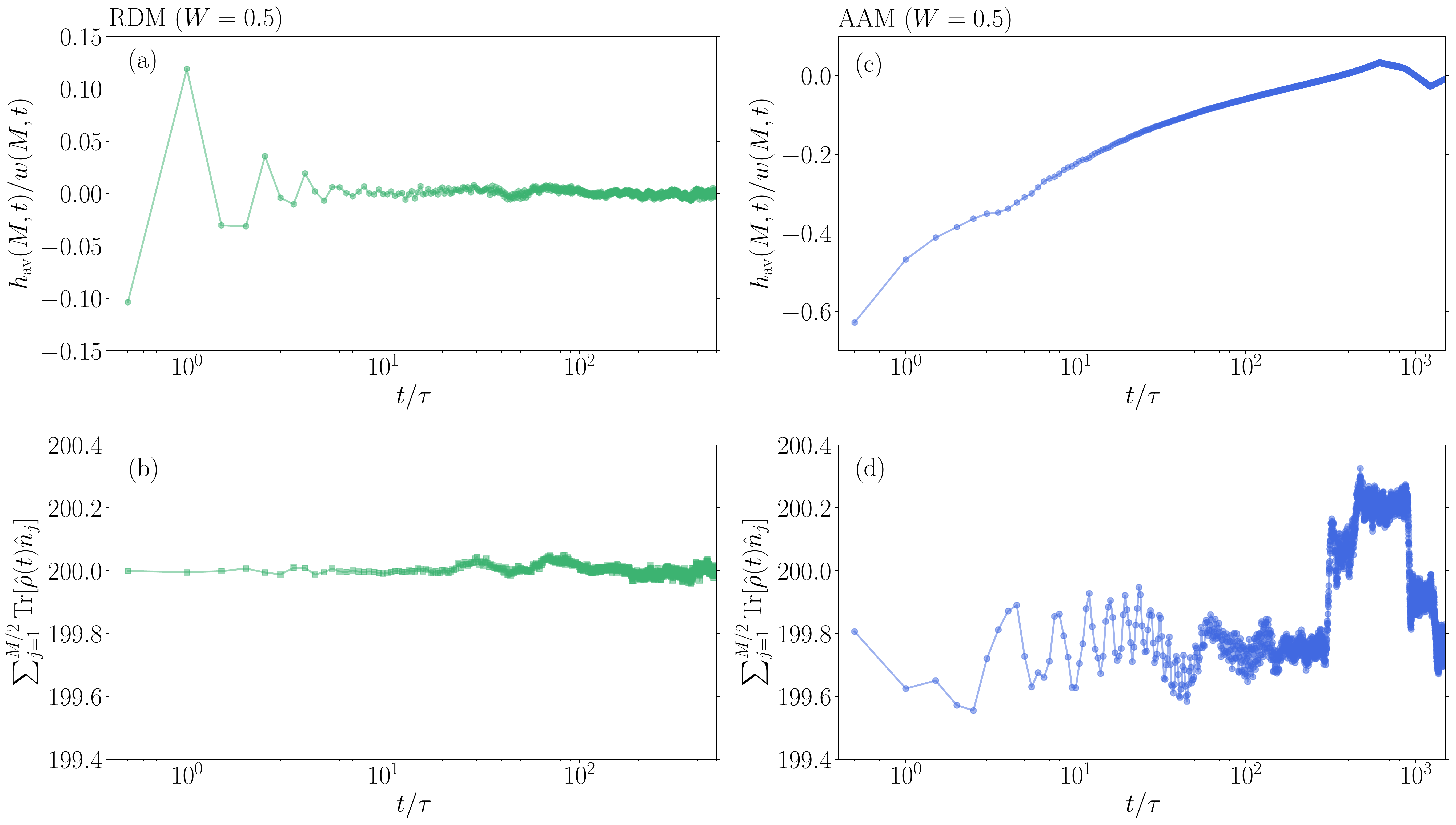}
\caption{
Numerical test of the assumptions (i) and (iii) for (a,b) the RDM and (c,d) the AAM with $W=0.5$ and $M=800$.  
(a,c) Time evolution of $h_{\rm av}(t)/w(M,t)$. We find that, in both of the models, the ratios become smaller than unity as time goes by, and then the assumption (i) works well. (b,d) Time evolution of $ \sum_{j=1}^{M/2} {\rm Tr} \left[ \hat{\rho}(t)  \hat{n}_{j} \right] $. One can see that the values are close to $N/2=M/4=200$, and thus the assumption (iii) is valid in the whole time regimes. 
} 
\label{validity_fig1} 
\end{center}
\end{figure}

\begin{figure}[t]
\begin{center}
\includegraphics[keepaspectratio, width=18cm,clip]{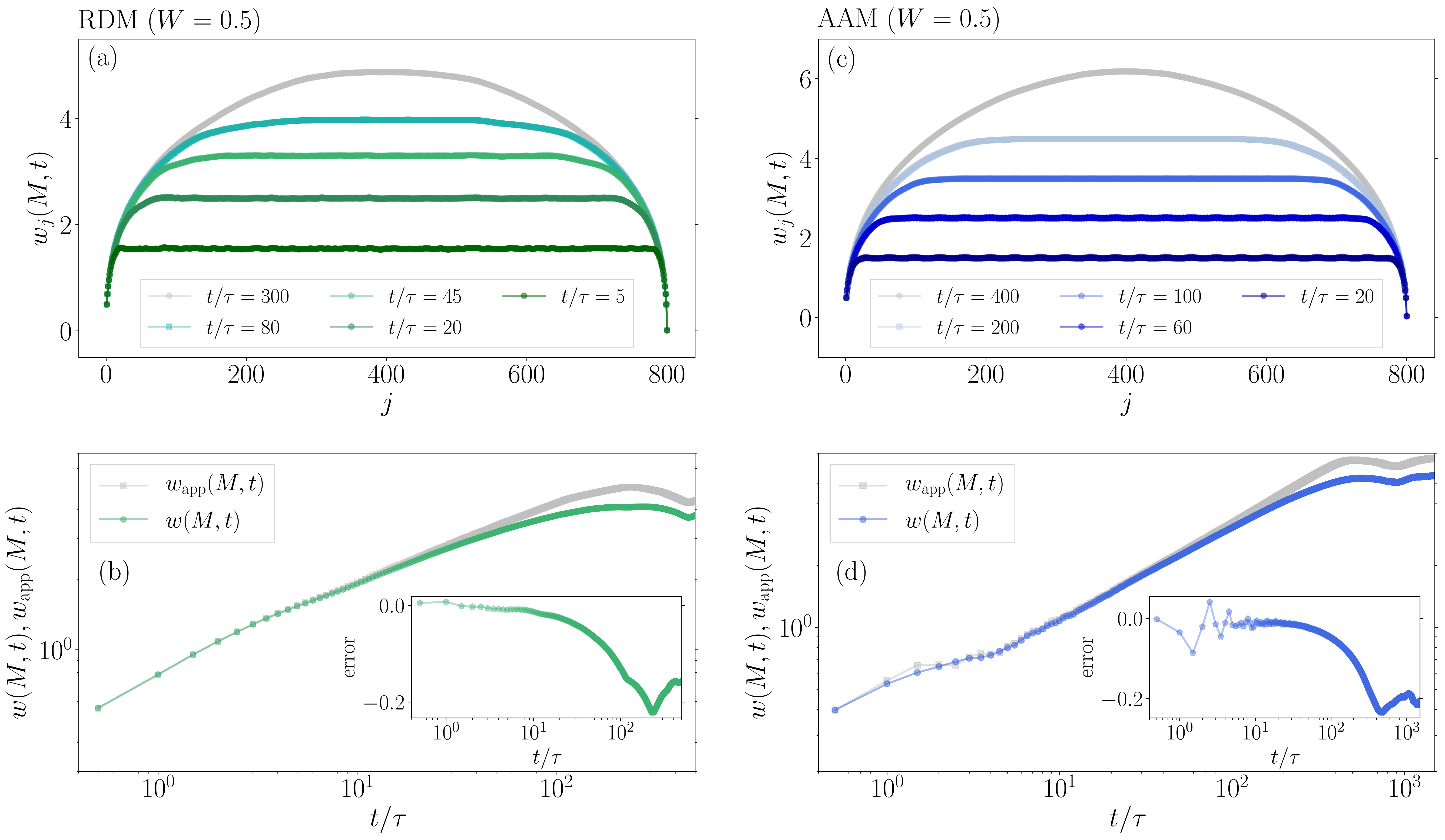}
\caption{
Numerical test of the assumption (ii) and Eq.~\eqref{roughness_ap22} in (a,b) the RDM and (c,d) the AAM with $W=0.5$ and $M=800$.  
(a,c) Time evolution of $w_{j}(M,t)$. In the early stage of the dynamics, the distribution is uniform far from the boundaries, but the uniformity breaks down in the late dynamics. 
(b,d) Time evolution of $w(M,t)$ and $w_{\rm app}(M,t)$. The insets plot the error $(w-w_{\rm app})/w$. The maximum deviation is about $-0.2$. 
} 
\label{validity_fig2} 
\end{center}
\end{figure}

Before applying the diagonal ensemble to the surface roughness, we first approximate the surface roughness by imposing the following assumption:
\begin{eqnarray}
&&{\rm Assumption~(i):}~h_{\rm av}(t) \simeq 0, \nonumber \\
\nonumber \\
&&{\rm Assumption~(ii):}~w(M,t)^2 \simeq  {\rm Tr} \left[ \hat{\rho}(t)  (\hat{h}_{M/2} - h_{\rm av}(t))^2  \right], \nonumber \\
&&{\rm Assumption~(iii):}~\sum_{j=1}^{M/2} {\rm Tr} \left[ \hat{\rho}(t)  \hat{n}_{j}  \right] \simeq \frac{\nu M}{2}. \nonumber 
\end{eqnarray}
While the first and second assumptions are difficult to prove rigorously, our numerical simulations confirm their validity as shown in Figs.~\ref{validity_fig1} and \ref{validity_fig2} as discussed later. The third assumption means that the particle number in the half of the system is equal to half of the total particle number, and we expect it to be valid in our system because the initial state is the staggered state and all the delocalized modes can spread over the whole system. Indeed, we numerically confirm the validity of assumption (iii) as shown in Fig.~\ref{validity_fig1}. We also note that assumption (iii) is analytically justified for the RDM after the average over disorder.

Under the assumptions (i) and (ii) and the definition of the surface-height operator, we obtain
\begin{eqnarray}
w(M,t)^2 &\simeq& {\rm Tr} \left[ \hat{\rho}(t)  \left(  \sum_{j=1}^{M/2} \hat{n}_j  - \frac{M \nu}{2}   \right) ^2  \right], \label{bipartite_w} \\ 
&=& \sum _{k,l=1}^{M/2}  {\rm Tr} \left[ \hat{\rho}(t)  (\hat{n}_k \hat{n}_l - \nu \hat{n}_k - \nu \hat{n}_l + \nu^2)   \right], \label{roughness_ap0}
\end{eqnarray}
It is worthy of mentioning here that Eq.~\eqref{bipartite_w} is equivalent to a square of the bipartite fluctuation~\cite{bipartite1,bipartite2,bipartite3}, which quantifies the particle-number fluctuation in the half of the system. 

Finally, applying the Wick decomposition~\eqref{correlation_M2} to Eq.~\eqref{roughness_ap0} and use the assumption (iii), we find
\begin{eqnarray}
w(M,t)^2 &\simeq& \sum_{j=1}^{M/2} D_{jj}(t)    - \sum_{i,j=1}^{M/2} D_{ij}(t) D_{ji}(t) + \left( \sum_{j=1}^{M/2} D_{jj}(t) \right)^2    - \frac{\nu^2 M^2}{4} \\
             &\simeq&  \sum_{j=1}^{M/2}  D_{jj}(t)     - \sum_{i,j=1}^{M/2}   D_{ij}(t) D_{ji}(t).
            \label{roughness_ap}
\end{eqnarray}
Here, to derive the last line, we utilize $\left( \sum_{j=1}^{M/2} D_{jj}(t) \right)^2 = \left( \sum_{j=1}^{M/2} {\rm Tr}  \left[ \hat{\rho}(t)  \hat{n}_{j}  \right]  \right)^2  = \nu^2 M^2/4$ owing to the assumption (iii). For the calculation in the following sections, we here define
\begin{eqnarray}
A(M,t) = \sum_{j=1}^{M/2}  D_{jj}(t), 
\label{roughness_ap1}
\end{eqnarray}
\begin{eqnarray}
B(M,t) =  \sum_{i,j=1}^{M/2}   D_{ij}(t) D_{ji}(t).
\label{roughness_ap2}
\end{eqnarray}
Then, we finally obtain the approximated surface roughness $w_{\rm app}(M,t)$, which is defined as
\begin{eqnarray}
w_{\rm app}(M,t)^2 := A(M,t) - B(M,t). 
 \label{roughness_ap22}
\end{eqnarray}

We numerically investigate the validity of the three assumptions given in the beginning of this section, and check whether or not Eq.~\eqref{roughness_ap} works well. Figure~\eqref{validity_fig1} shows the time evolution for $h_{\rm av}(t)/w(M,t)$ and $ \sum_{j=1}^{M/2} {\rm Tr} \left[ \hat{\rho}(t)  \hat{n}_{j} \right] $, from which we find that the assumptions (i) and (iii) are valid. Note that, in the AAM, the averaged surface-height is not much smaller than unity but becomes smaller as time goes by. 

In the upper panels of Fig.~\ref{validity_fig2}, in order to consider the assumption (ii), we plot the site-dependent surface roughness defined by
\begin{eqnarray}
w_{j}(M,t) = \left[ \hat{\rho}(t)  (\hat{h}_{j} - h_{\rm av}(t))^2  \right].
\end{eqnarray}
For both of the AAM and the RDM, $w_{j}(M,t)$ in the early stage of the dynamics is almost independent of the site $j$ except for the edges. 
The non-uniformity, however, appears even at the center as time goes by, and thus the assumption (ii) becomes worse in the late stage. 
Actually, as shown in the lower panels of Fig.~\ref{validity_fig2}, $w_{\rm app}(M,t)$ begins to deviate from $w(M,t)$ in $t> 60 \tau$ and $t > 200\tau$ for the RDM and the AAM, respectively. Note that, as described in Fig.~4 of the main text, the FV-type scaling of the EE still holds.

\subsection{$w_{\rm app}(M,t)$ on the diagonal ensemble}
We apply the diagonal ensemble to Eq.~\eqref{roughness_ap22}.  
For this purpose, we first consider time dependence of the correlation matrix, which is determined by the eigenvalues $\epsilon_{\alpha}~(\alpha=1,\cdots,M)$. 
Using the operators for the quasi-particles, we can obtain
\begin{eqnarray}
D_{ij}(t) &=& {\rm Tr} \left[ \hat{\rho}(0) \hat{f}_i^{\dagger}(t) \hat{f}_j(t) \right] \\
&=& \sum_{\alpha, \beta=1}^{M}  v_{i \alpha}^* v_{j \beta}  {\rm Tr} \left[ \hat{\rho}(0) \hat{F}_{\alpha}^{\dagger}(t) \hat{F}_{\beta}(t) \right] \\
&=& \sum_{\alpha, \beta=1}^{M}  v_{i \alpha}^* v_{j \beta}  {\rm Tr} \left[ \hat{\rho}(0) \hat{F}_{\alpha}^{\dagger} \hat{F}_{\beta} \right] e^{ i( \epsilon_{\alpha} - \epsilon_{\beta} )t/\hbar}.
\label{D_exact}
\end{eqnarray}
Then, substituting the exact result~\eqref{D_exact} into $A(t)$ and taking the long time average by assuming no degeneracy, we obtain 
\begin{eqnarray}
A_{\rm ave} &:=& \lim_{T \to \infty} \frac{1}{T} \int_{0}^{T} dt A(t) \\
&=& \lim_{T \to \infty} \frac{1}{T} \int_{0}^{T} dt  \sum_{j=1}^{M/2} D_{jj}(t) \\
&=& \sum_{j=1}^{M/2} \sum_{\alpha, \beta=1}^{M}  v_{j \alpha}^* v_{j \beta}  {\rm Tr} \left[ \hat{\rho}(0) \hat{F}_{\alpha}^{\dagger} \hat{F}_{\beta} \right] 
\delta_{\alpha \beta} \\
&=& \sum_{j=1}^{M/2} \sum_{\alpha=1}^{M}   v_{j \alpha}^* v_{j \alpha}  {\rm Tr} \left[ \hat{\rho}(0)  \hat{F}_{\alpha}^{\dagger} \hat{F}_{\alpha}  \right].
\end{eqnarray}
Defining $I_{\alpha \beta} = \sum_{j=1}^{M/2} v_{j \alpha}^* v_{j \beta}$, we obtain
\begin{eqnarray}
A_{\rm ave} = \sum_{\alpha=1}^{M}  I_{\alpha \alpha}   {\rm Tr} \left[ \hat{\rho}(0) \hat{F}_{\alpha}^{\dagger} \hat{F}_{\alpha} \right]  . 
\label{A_ave}
\end{eqnarray}
Following the same way, we calculate the time-averaged value of $B(t)$ as
\begin{eqnarray}
B_{\rm ave} &:=& \lim_{T \to \infty} \frac{1}{T} \int_{0}^{T} dt B(t) \\
&=& \sum_{\alpha, \beta, \mu, \nu=1}^{M}    I_{\alpha \nu} I_{\mu \beta}   
{\rm Tr} \left[ \hat{\rho}(0) \hat{F}_{\alpha}^{\dagger} \hat{F}_{\beta} \right]
{\rm Tr} \left[ \hat{\rho}(0) \hat{F}_{\mu}^{\dagger} \hat{F}_{\nu} \right]  \delta_{\alpha + \mu, \beta + \nu }. 
\label{B_ave}
\end{eqnarray}
Thus, the substitution of Eqs.~\eqref{A_ave} and \eqref{B_ave} into Eq.~\eqref{roughness_ap22} leads to the surface~roughness in the stationary state:
\begin{eqnarray}
w_{\rm ave}(M)^2 &:=& \lim_{T \to \infty} \frac{1}{T} \int_{0}^{T} dt w_{\rm app}(M,t)^2 \\
&=&  A_{\rm ave} - B_{\rm ave} . 
\label{roughness_ap3}
\end{eqnarray}
Thus, instead of directly solving the time-dependent Schrödinger equation, we can calculate the surface-roughness in the stationary state.  

\begin{figure}[t]
\begin{center}
\includegraphics[keepaspectratio, width=18cm,clip]{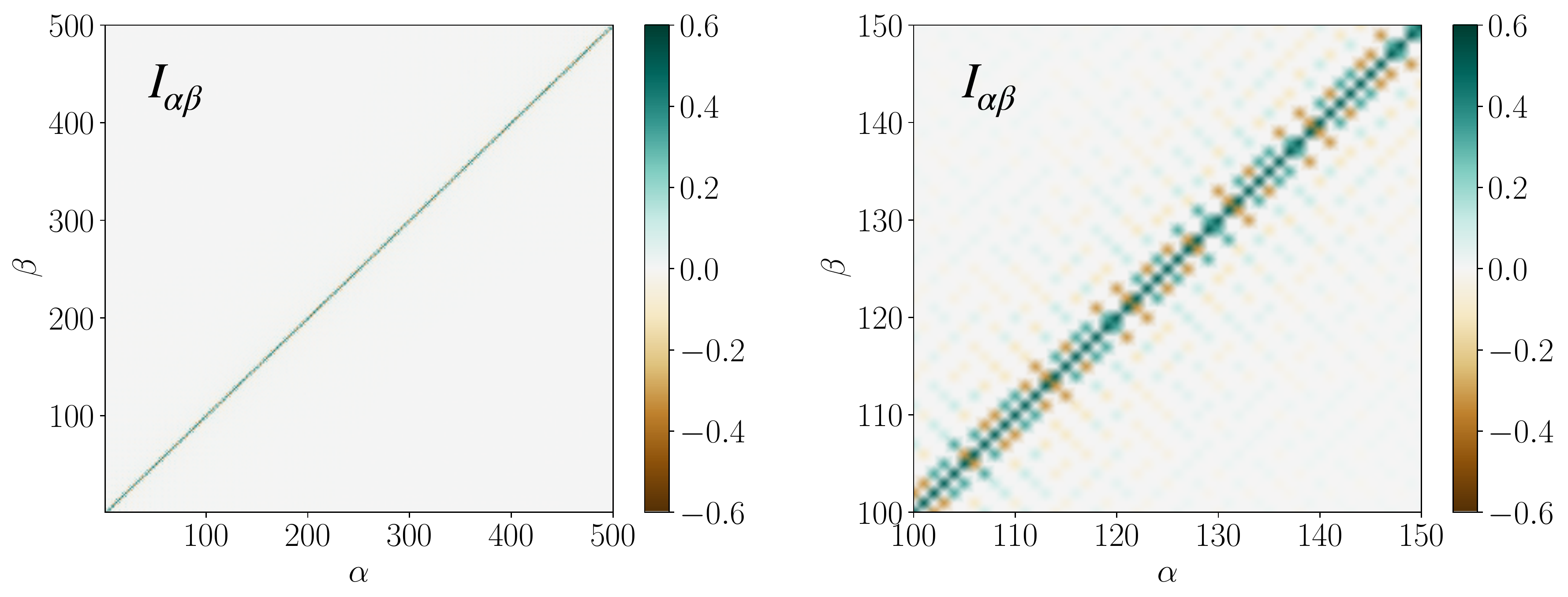}
\caption{
Numerical result of $I_{\alpha \beta}$ in the non-interacting fermion with $V_j=0$ and $M=500$. The left panel shows all the data, and the right one is the enlarged figure. 
} 
\label{I_graph_sup_fig} 
\end{center}
\end{figure}

\subsubsection{Derivation of the FV scaling exponent $\alpha=0.5$ in a fermion system free from disorder. }
Before discussing disordered systems, we show that normal exponent $\alpha=1/2$ is obtained for several systems whose eigenstates are all delocalized. First, we consider a fermion system with $V_j=0$ by using Eq.~\eqref{roughness_ap3}. 
In this model, the eigenfunctions $v_{j \alpha}$ are expressed by the plane waves: 
\begin{eqnarray}
v_{j 0} &=& \frac{1}{\sqrt{M}} \\
v_{j M} &=& \frac{1}{\sqrt{M}} (-1)^j \\
v_{j \alpha}^{\rm even} &=& \sqrt{\frac{2}{M}} \cos \left(  2 \pi \alpha j/M \right) ~~~~~(\alpha = 1, 2,\cdots, M/2-1), \\
v_{j \alpha}^{\rm odd}  &=& \sqrt{\frac{2}{M}} \sin \left(  2 \pi \alpha j/M \right)~~~~(\alpha = 1, 2, \cdots, M/2-1),
\label{Derivation_alpha1}
\end{eqnarray}
where the corresponding eigenvalue is given by $\epsilon_{\alpha} = -2J \cos(2 \pi \alpha/M)$ with the integer label $\alpha$.
Using this expression, we can evaluate $I_{\alpha \beta} = \sum_{j=1}^{M/2} v_{j \alpha}^* v_{j \beta}$, but do not show the concrete expression since they are too complicated. 
As shown in Fig.~\ref{I_graph_sup_fig}, in large $M$, $I_{\alpha \beta}$ is well approximated to be
\begin{eqnarray}
I_{\alpha \beta} \simeq \frac{1}{2} \delta_{\alpha \beta}.
\label{Derivation_alpha12}
\end{eqnarray} 
We comment on negative and positive values of the off-diagonal component $I_{\alpha \beta}$ in Fig.~\ref{I_graph_sup_fig}. 
Their contribution in Eqs.~\eqref{A_ave} and \eqref{B_ave} will be small because the summation leads to the cancelation. 

We substitute Eq.~\eqref{Derivation_alpha12} into Eqs.~\eqref{A_ave} and \eqref{B_ave}, and then obtain
\begin{eqnarray}
A_{\rm ave} \simeq \frac{1}{2} \sum_{\alpha=1}^{M}  {\rm Tr} \left[ \hat{\rho}(0) \hat{F}_{\alpha}^{\dagger} \hat{F}_{\alpha} \right]. 
\label{Derivation_alpha2}
\end{eqnarray}
\begin{eqnarray}
B_{\rm ave} \simeq \frac{1}{4} \sum_{\alpha, \beta=1}^{M}  \left| {\rm Tr} \left[ \hat{\rho}(0) \hat{F}_{\alpha}^{\dagger} \hat{F}_{\beta} \right] \right|^2.
\label{Derivation_alpha3}
\end{eqnarray}
Next, using Eqs.~\eqref{quasi1}, \eqref{quasi2}, and \eqref{Derivation_alpha1}, we evaluate ${\rm Tr} \left[ \hat{\rho}(0) \hat{F}_{\alpha}^{\dagger} \hat{F}_{\beta} \right]$ as
\begin{eqnarray}
{\rm Tr} \left[ \hat{\rho}(0) \hat{F}_{\alpha}^{\dagger} \hat{F}_{\beta} \right] 
&=& \sum_{i=1}^{M} \sum_{j=1}^{M} v_{i \alpha} v_{j \beta}^* {\rm Tr} \left[ \hat{\rho}(0) \hat{f}_{i}^{\dagger} \hat{f}_{j} \right] \\
&=& \sum_{j=1}^{M/2} v_{(2j) \alpha} v_{(2j) \beta}^*
\label{Derivation_alpha4}
\end{eqnarray}
Substituting Eq.~\eqref{Derivation_alpha4} into Eqs.~\eqref{Derivation_alpha2} and \eqref{Derivation_alpha3}, we derive
\begin{eqnarray}
A_{\rm ave} &\simeq& \frac{M}{4}, 
\label{Derivation_alpha5} \\
B_{\rm ave} &\simeq& \frac{M}{8}
\label{Derivation_alpha6}
\end{eqnarray}
Therefore, we finally obtain 
\begin{eqnarray}
w_{\rm ave}(M)^2 \simeq \frac{1}{8}M.
\label{Derivation_alpha7}
\end{eqnarray}
Note that the degeneracy of the eigenvalue $\epsilon_{\alpha}$ exists, which seems to be inconsistent with the assumption for the diagonal ensemble. However, Eqs.~\eqref{A_ave} and \eqref{B_ave} still hold because of Eq.~\eqref{Derivation_alpha12}. Actually, only by using Eq.~\eqref{Derivation_alpha12}, we can directly derive Eqs.~\eqref{A_ave} and \eqref{B_ave} without utilizing the non-degeneracy assumption.   

\begin{figure}[t]
\begin{center}
\includegraphics[keepaspectratio, width=18cm,clip]{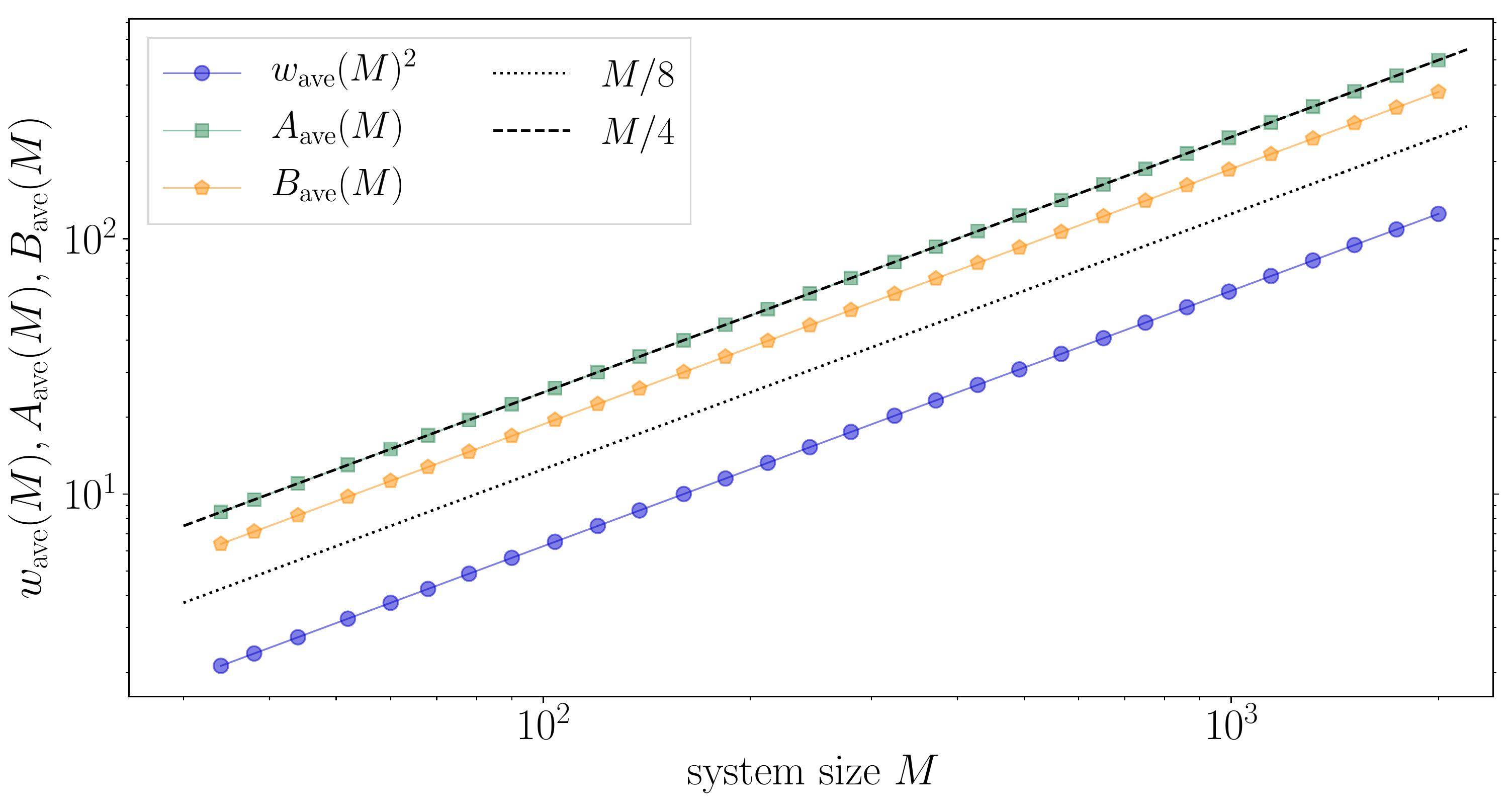}
\caption{
Numerical test of Eqs.~\eqref{Derivation_alpha5}, \eqref{Derivation_alpha6}, and \eqref{Derivation_alpha7}.
Solving the eigenvalue problem~\eqref{App1} with $V_j=0$ under the periodic boundary condition, we calculate Eqs.~\eqref{A_ave}, \eqref{B_ave}, and \eqref{roughness_ap3}. One can see that Eqs.~\eqref{Derivation_alpha5} shows the excellent agreement with the numerical results, while Eqs.~\eqref{Derivation_alpha6} and \eqref{Derivation_alpha7} deviates from the numerical ones but correctly reproduce the $M$-power-law dependence. This deviation is attributed to the approximation of Eq.~\eqref{Derivation_alpha12}. 
} 
\label{app_roughness_sup_fig} 
\end{center}
\end{figure}

We check the validity of Eqs.~\eqref{Derivation_alpha5}, \eqref{Derivation_alpha6}, and \eqref{Derivation_alpha7} by numerically calculating Eqs.~\eqref{A_ave}, \eqref{B_ave}, and \eqref{roughness_ap3}. Figure~\eqref{app_roughness_sup_fig} shows that Eqs.~\eqref{Derivation_alpha5} works well, while Eqs.~\eqref{Derivation_alpha6} and \eqref{Derivation_alpha7} does not.
This deviation comes from the approximation of Eq.~\eqref{Derivation_alpha12}. However, all the analytical results correctly reproduce the $M$-power law, and thus the free-fermion system free from disorder potentials have a normal exponent $\alpha=1/2$. Moreover, as discussed in Ref.~\cite{height_op2}, systems satisfying the eigenstate thermalization hypothesis also have $\alpha=1/2$. Therefore, we can conclude that generic delocalized systems have $\alpha=1/2$. 

\subsubsection{Numerical results for the AAM and the RDM}
Next, we consider disordered systems. 
Figure 3 of the main text shows the dependence of $w_{\rm ave} $ on the system size $M$, which is numerically obtained by diagonalizing the matrix $H$ of Eq.~\eqref{Hmat} and using Eqs.~\eqref{roughness_ap3}. We find that the stationary surface-roughness in the RDM and the AAM scales as $M^{0.33}$ and $M^{0.5}$, which are consistent with the exponents $\alpha=0.33$ and $0.5$ in the FV scaling discussed in the main text. Thus, we can reproduce the anomalous exponent using only eigenstates and the initial state.

Remember that $\alpha=1/2$ is obtained for certain systems with completely delocalized states, i.e., the fermion system free from disorder and systems obeying the eigenstate thermalization hypothesis. Thus, we argue that the partially LESs in the delocalized phase of the RDM play a significant role in the emergence of the anomalous exponent $\alpha=0.33$.

\clearpage

\section{Relation between the von Neumann entanglement entropy and the surface roughness}
In the main text, we investigate the entanglement dynamics and its dynamical one-parameter scaling by considering the relation between the von Neumann entanglement entropy (EE) and the surface roughness. Here, we describe how to derive the relation and a related useful inequality.

\subsection{Relation between $S_{\rm EE}(M,t)$ and $w(M,t)$}
The EE in our models can be expressed by the eigenvalues of the correlation matrix $D_{ij}$.
This expression is useful to derive the relation between $S_{\rm EE}(M,t)$ and $w(M,t)$.  
In the main text, the EE is defined by
\begin{eqnarray}
S_{\rm EE}(M,t) = - {\rm Tr}' \left[  \hat{\rho}_{\rm re}(t) \log \hat{\rho}_{\rm re}(t) \right], 
\end{eqnarray}
where a reduced density matrix is $\hat{\rho}_{\rm re}(t) = {\rm tr}_{B} \hat{\rho}_{\rm pure}(t)$ with the set $B = \{ M/2 + 1, \cdots, M  \}$ and ${\rm Tr}'$ denotes a partial trace excluding $B$. Here, $\hat{\rho}_{\rm pure}(t)$ is a density matrix for a single realization, and we take the ensemble average for $S_{\rm EE}(M,t)$ in the RDM. According to the previous work~\cite{Cao2019}, $S_{\rm EE}(M,t)$ becomes
\begin{eqnarray}
S_{\rm EE}(M,t) = - \sum_{n=1}^{M/2}   \lambda_{n}(t) \log \lambda_{n}(t)  - \sum_{n=1}^{M/2}    \left( 1-\lambda_{n}(t) \right) \log \left(1-\lambda_{n}(t) \right),
\label{ee1}
\end{eqnarray}
where $\lambda_{n}(t) \in [0,1]~(n=1,\cdots,M/2)$ are eigenvalues of the correlation matrix $D_{ij}(t)~(i,j=1,\cdots,M/2)$.  
On the other hand, the approximated surface roughness~\eqref{roughness_ap} can be expressed by the eigenvalues:
\begin{eqnarray}
w(M,t)^2 &\simeq& \sum_{n=1}^{M/2} \lambda_n(t) \left( 1 - \lambda_n(t) \right) \\
& = & w_{\rm app}(M,t) ^2~~~~(\because \eqref{roughness_ap22}).
\label{roughness_ap2}
\end{eqnarray}

\begin{figure}[t]
\begin{center}
\includegraphics[keepaspectratio, width=18cm,clip]{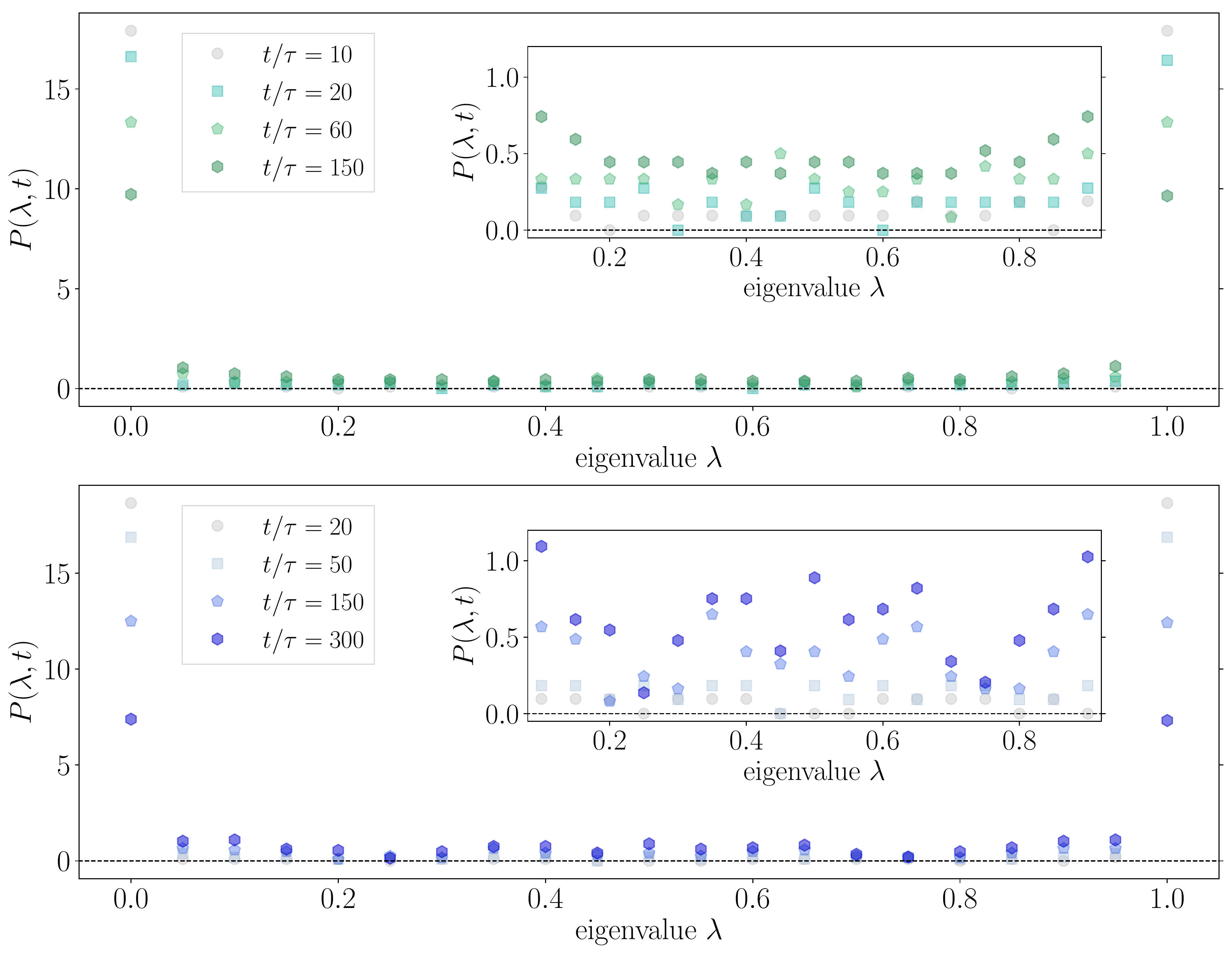}
\caption{Probability density $P(\lambda,t)$ for the eigenvalues $\lambda_n(t)$ of the correlation matrix $D_{ij}(t)~(i,j=1,\cdots,M/2)$. 
The upper and lower panels show the numerical results for the RDM and the AAM with $W=0.5$ and $M=800$, respectively.
The insets are the enlarged figures for $\lambda \in [0.08,0.92]$
} 
\label{probability_fig} 
\end{center}
\end{figure}

To connect $S_{\rm EE}(M,t)$ with $w(M,t)$, we introduce the probability function $P(\lambda,t)$ for $\lambda_n(t)$. 
Then, we can rewrite $S_{\rm EE}(M,t)$ and $w(M,t)^2$ as
\begin{eqnarray}
S_{\rm EE}(M,t) &=& -  \int d \lambda P(\lambda,t)  \left\{  \lambda \log \lambda  +  \left( 1-\lambda \right) \log \left(1-\lambda \right) \right\}, \label{ee_hosoku}
\\
w(M,t)^2 &\simeq&   \int d \lambda P(\lambda,t)   \lambda (1 - \lambda).
\label{roughness_ap4}
\end{eqnarray}
Here, we assume 
\begin{eqnarray}
P(\lambda,t) \simeq p_1(t) \delta(\lambda) + p_2(t) \theta(1-\lambda) \theta(\lambda) + p_3(t) \delta(\lambda-1), 
\end{eqnarray}
where $p_j(t)~(j=1,2,3)$ is a time-dependent weight independent of $\lambda$ and $\theta(\cdot)$ is the Heaviside step function. 
The crucial assumption given here is that the probability distribution is uniform for $0<\lambda<1$.
As shown in Fig.~\ref{probability_fig}, the assumption is well satisfied in both the RDM and the AAM. 
Then, using this assumption, we get
\begin{eqnarray}
S_{\rm EE}(M,t) &\simeq &-  p_2(t) \int_{0}^1 d \lambda  \left\{  \lambda \log \lambda  +  \left( 1-\lambda \right) \log \left(1-\lambda \right) \right\}, 
\label{ee_hosoku1}
\\
w(M,t)^2 &\simeq&    p_2(t) \int_0^1 d \lambda  \lambda (1 - \lambda).
\label{roughness_ap41}
\end{eqnarray}
Finally, we note the following integral formula:
\begin{eqnarray}
- \int_0^1 ( x \log x  + (1-x) \log (1-x) ) dx = \int_0^1 3x(1-x) dx. 
\label{integral1}
\end{eqnarray}
We use the formula in  Eqs.~\eqref{ee_hosoku1} and \eqref{roughness_ap41}, which leads to
\begin{eqnarray}
S_{\rm EE}(M,t) \simeq 3 w(M,t)^2.
\label{ee3}
\end{eqnarray}
The insets of Fig.~4 in the main text show that Eq.~\eqref{ee3} works well especially in the early stage of the dynamics. 
In the late stage, the relation becomes a little worse because Eq.~\eqref{roughness_ap2} becomes worse, but the roughness still captures the qualitative behavior of the von Neumann EE.

\begin{figure}[t]
\begin{center}
\includegraphics[keepaspectratio, width=18cm,clip]{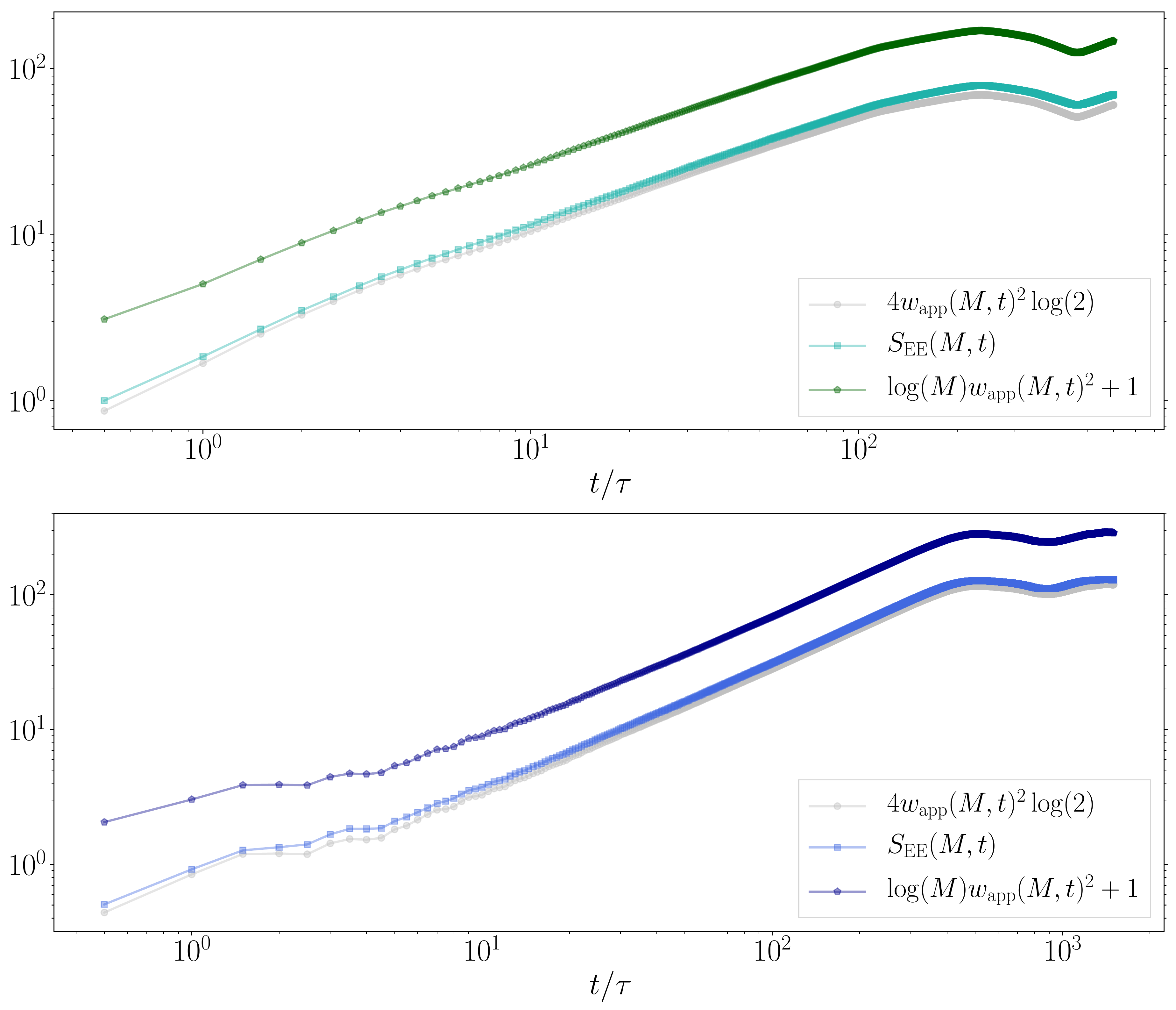}
\caption{Numerical verification of the inequality~\eqref{inequ}. The upper and lower panels show the results for the RDM and the AAM with $W=0.5$ and $M=800$, respectively.} 
\label{inequality_fig} 
\end{center}
\end{figure}

\subsection{Power-law growth of $S_{\rm EE}(M,t)$ and $w_{\rm app}(M,t)$}
Our numerical results demonstrate that the EE $S_{\rm EE}(M,t)$ and the approximated surface-roughness $w_{\rm app}(M,t)$ obey the power-law growth.
We here derive the relation between the two power exponents in the thermodynamic limit. 

\subsubsection{Inequlity of $S_{\rm EE}(M,t)$ and $w_{\rm app}(M,t)$}
We first prove a useful inequality for $S_{\rm EE}(t,M)$. 
The approximated surface-roughness $w_{\rm app}(M,t)$ is expressed by
\begin{eqnarray}
w_{\rm app}(M,t)^2 = \sum_{n=1}^{M/2} \lambda_n(t) \left( 1 - \lambda_n(t) \right).
\label{roughness_ap5}
\end{eqnarray}

We note that the following inequality is derived for $x \in [0,1]$ and $M>15$:
\begin{eqnarray}
4 x (1-x) \log (2) < - x \log (x) - (1-x) \log (1-x) < x (1-x) \log(M) + \frac{2}{M}. 
\label{ee4}
\end{eqnarray}
Finally, by using Eqs.~\eqref{ee1}, \eqref{roughness_ap5}, and \eqref{ee4}, we obtain 
\begin{eqnarray}
4 w_{\rm app}(M,t)^2 \log (2) < S_{\rm EE}(M,t) <  w_{\rm app}(M,t)^2 \log(M)  + 1.
\label{inequ}
\end{eqnarray}
Figure~\ref{inequality_fig} numerically checks Eq.~\eqref{inequ} in the RDM and the AAM.

\subsubsection{Power exponents in the thermodynamic limit}
According to our numerical results, the surface roughness shows the FV scaling characterized by the universal exponents $(\alpha,\beta,z) > \bm{0}$  for sufficiently large systems. Then, it is reasonable to assume that, for any positive and a real number $\epsilon$, there exists an integer $M_0$ such that the surface roughness satisfies
\begin{eqnarray}
\left| \frac{ \log w_{\rm app}(M,t)}{\log t} - \beta \right| < \epsilon
\label{sup:thermo1}
\end{eqnarray}
for any $M>M_0$ and $aM^{u_1} < t < b M^{u_2}$ with given positive constants $a$, $b$, and $u_1<u_2<z$.
Under this assumption, we can prove the following proposition.

\begin{prop}  
For any positive and a real number $\delta$, there exists an integer $M_1$ such that the EE satisfies
\begin{eqnarray}
\left| \frac{ \log S_{EE}(M,t)}{\log t} - 2\beta \right| < \delta
\end{eqnarray}
for any $M>M_1$ and $aM^{u_1} < t < b M^{u_2}$. 
\end{prop} 
\noindent\textit{Proof.}
The assumption~\eqref{sup:thermo1} leads to 
\begin{eqnarray}
t^{\beta - \epsilon} < w_{\rm app}(M,t) < t^{\beta + \epsilon}.
\label{sup:thermo2}
\end{eqnarray}
Substituting $t=aM^{u_1}$ into the lower bound of Eq.~\eqref{sup:thermo2}, we obtain
\begin{eqnarray}
a^{\beta - \epsilon} M^{(\beta - \epsilon)u_1} < w_{\rm app}(M,t), 
\label{sup:thermo3}
\end{eqnarray}
from which we can always take an integer $M_2~(>M_0)$ ensuring $1<w_{\rm app}(M,t) $ and $0<\log t$ for $M > M_2$ because the left-hand side of Eq.~\eqref{sup:thermo3} and the lower bound $aM^{u_1}$ of the time regime increase with $M$. 

Using the inequality of Eq.~\eqref{inequ}, we derive
\begin{eqnarray}
\frac{2 \log w_{\rm app}(M,t) + \log (4\log2 )}{\log t}   <   \frac{ \log S_{EE}(M,t)}{\log t}   <    \frac{ \log( w_{\rm app}(M,t)^2 (\log M)   + 1 ) }{\log t} .
\label{sup:thermo4}
\end{eqnarray}
The fact $1<w_{\rm app}(M,t) $ leads to 
\begin{eqnarray}
 \frac{ \log( w_{\rm app}(M,t)^2 (\log M)   + 1 ) }{\log t}   <   \frac{ 2\log w_{\rm app}(M,t) + \log(\log(M) + 1)  }{\log t}  .
\label{sup:thermo4}
\end{eqnarray}
Then, for $M>M_2$ we obtain 
\begin{eqnarray}
\frac{2 \log w_{\rm app}(M,t) + \log (4\log2 )}{\log t}   <   \frac{ \log S_{EE}(M,t)}{\log t}   <    \frac{ 2\log w_{\rm app}(M,t) + \log(\log(M) + 1)  }{\log t}  . 
\label{sup:thermo5}
\end{eqnarray}
Using Eqs.~\eqref{sup:thermo2} and \eqref{sup:thermo5}, we obtain
\begin{eqnarray}
 -2 \epsilon +  \frac{\log (4\log2 )}{\log t}   <   \frac{ \log S_{EE}(M,t)}{\log t} - 2\beta  <   2  \epsilon + \frac{  \log(\log(M) + 1)  }{\log t} . 
\label{sup:thermo6}
\end{eqnarray}
For any positive and a real number $\delta$, considering the condition $aM^{u_1} < t < b M^{u_2}$ and setting $\epsilon = \delta/4$, we can always take an integer $M_3$ such that
\begin{eqnarray}
{\rm max} \left(  \left| -2 \epsilon +  \frac{\log (4\log2 )}{\log t}  \right| , \left| 2  \epsilon + \frac{  \log(\log(M) + 1)  }{\log t}  \right| \right) < \delta .
\end{eqnarray}
for $M>M_3$.
Thus, we finally obtain 
\begin{eqnarray}
\left| \frac{ \log S_{EE}(M,t)}{\log t} - 2\beta  \right| <   \delta
\label{sup:thermo6}
\end{eqnarray}
for $M> {\rm max} (M_2,M_3) := M_1$. This completes the proof. \sq
\\
\\

This proposition means that the EE grows with $t^{2  \beta}$ in the thermodynamic limit if the approximated surface-roughness (which is also approximated by the bipartite particle fluctuation) grows with $t^{\beta}$.

\section{Anomalous behavior of single-particle transport in the RDM}
We discuss that the RDM exhibits anomalous behavior in single-particle transport properties as well as the FV scaling with the anomalous exponents. As already mentioned, in the RDM with $W<1$, the number of the DLESs is proportional to $\sqrt{M}$. Owing to this $\sqrt{M}$ dependence, the standard deviation $\Delta x^2$ of the position of a particle initially localized at a certain site grows as $t^{3/2}$ \cite{RDM1}, differently from $t^2$ growth in a disorder-free system. Similarly, the saturated value of $\Delta x$ at late time is found to be proportional to $M^{3/4}$ rather than $M$ as shown below. 
From these results, one might expect $z=4/3$, but it is different from the dynamical FV scaling exponent $z=1.00$. 
It is an interesting and nontrivial future problem to investigate whether these single-particle anomalous exponents are related to the FV scaling.

In what follows, we explain the fact that the saturated $\Delta x$ is proportional to $M^{3/4}$.
We consider a single-particle dynamics starting from the following state:
\begin{eqnarray}
\ket{\psi_{\rm single}} = \hat{f}^{\dagger}_{M/2} \ket{0}. 
\end{eqnarray}
Using the Schrödinger equation with this initial state, we calculate the deviation $\triangle x (t)$ from the center defined by
\begin{eqnarray}
\triangle x (t) ^2 = \sum_{j=1}^{M} \left( j -\frac{M}{2} \right)^2 \bra{\psi_{\rm single}} \hat{f}^{\dagger}_j (t) \hat{f}_{j} (t)  \ket{\psi_{\rm single}}. 
\end{eqnarray}
In the quasi-particle representation with Eqs.~\eqref{quasi1} and \eqref{quasi2}, this is expressed by
\begin{eqnarray}
\triangle x (t) ^2 = \sum_{j=1}^{M} \sum_{\alpha=1}^{M} \sum_{\beta=1}^{M} \left( j -\frac{M}{2} \right)^2    u_{j \alpha}^* u_{j \beta}  \nonumber \\
 \times \bra{\psi_{\rm single}} \hat{F}^{\dagger}_{\alpha}\hat{F}_{\beta}  \ket{\psi_{\rm single}} e^{\frac{i}{\hbar} \left(  \epsilon_{\alpha} - \epsilon_{\beta} \right)t}.  
\end{eqnarray}
Just as the calculation of the diagonal ensemble, we apply the long-time average, and then obtain the stationary deviation:
\begin{eqnarray}
\triangle x _{\rm stat} ^2 &:=& \lim_{T \rightarrow \infty} \frac{1}{T} \int_{0}^{T} \triangle x (t_1) ^2 dt_1  \\
&=& \sum_{j=1}^{M} \sum_{\alpha=1}^{M} \left( j -\frac{M}{2} \right)^2   |u_{j \alpha}|^2 \bra{\psi_{\rm single}} \hat{F}^{\dagger}_{\alpha}\hat{F}_{\alpha}  \ket{\psi_{\rm single}}. \nonumber \\
\end{eqnarray}
Here, using the initial state, we derive
\begin{eqnarray}
 \bra{\psi_{\rm single}} \hat{F}^{\dagger}_{\alpha}\hat{F}_{\alpha}  \ket{\psi_{\rm single}} = |u_{M/2, \alpha}|^2.
\end{eqnarray}
Thus, this leads to 
\begin{eqnarray}
\triangle x _{\rm stat} ^2 = \sum_{j=1}^{M} \sum_{\alpha=1}^{M} \left( j -\frac{M}{2} \right)^2   |u_{j \alpha}|^2  |u_{M/2, \alpha}|^2. \label{devi_stat1}
\end{eqnarray}

We estimate Eq.~\eqref{devi_stat1} by noting the fact that the RDM with $W<1$ has localized and delocalized eigenstates, and their numbers are proportional to $M$ and $\sqrt{M}$, respectively. Let us denote a set of labels $\alpha$ for the localized (delocalized) states  by $\mathcal{L}$ ($\mathcal{D}$). This notation gives
\begin{eqnarray}
\triangle x _{\rm stat} ^2 = && \sum_{j=1}^{M} \sum_{\alpha \in \mathcal{L}} \left( j -\frac{M}{2} \right)^2   |u_{j \alpha}|^2  |u_{M/2, \alpha}|^2 + \sum_{j=1}^{M} \sum_{\alpha \in \mathcal{D}} \left( j -\frac{M}{2} \right)^2   |u_{j \alpha}|^2  |u_{M/2, \alpha}|^2. 
\label{devi_stat2}
\end{eqnarray}

In the first term on the right hand side of Eq.~\eqref{devi_stat2}, the product of the eigenfunctions has large values around $j=M/2$ and $\alpha=\alpha_{M/2}$ whose eigenstates are spatially localized around $j=M/2$. We expect that the number of the $j$ and $\alpha$-summation does not increase with $M$. As a result, we obtain
\begin{eqnarray}
\triangle x _{\rm stat} ^2 =  C_{\rm loc} + \sum_{j=1}^{M} \sum_{\alpha \in \mathcal{D}} \left( j -\frac{M}{2} \right)^2   |u_{j \alpha}|^2  |u_{M/2, \alpha}|^2
\label{devi_stat3}
\end{eqnarray}
with a constant $C_{\rm loc} $ that does not increase with $M$. On the other hand, in the second term, we estimate 
\begin{eqnarray}
|u_{j \alpha}|^2  \sim  \frac{1}{M},
\end{eqnarray}
\begin{eqnarray}
|u_{M/2, \alpha}|^2 \sim \frac{1}{M}
\end{eqnarray}
because these states are delocalized. Thus, the stationary deviation becomes
\begin{eqnarray}
\triangle x _{\rm stat} ^2 
&=&  C_{\rm loc} +  \frac{1}{M^2} \sum_{j=1}^{M} \sum_{\alpha \in \mathcal{D}} \left( j -\frac{M}{2} \right)^2 \\
&\simeq&  C_{\rm loc} +  \frac{ C_{\rm deloc} \sqrt{M} }{M^2}  \sum_{j=1}^{M} \left( j -\frac{M}{2} \right)^2 \label{devi_stat4_hosoku}\\
&\simeq&  C_{\rm loc} +  \frac{C_{\rm deloc} \sqrt{M} }{3 M^2} M^3 + \cdots \\
&=&  C_{\rm loc} +  \frac{C_{\rm deloc}}{3}M^{3/2} + \cdots
\label{devi_stat4}
\end{eqnarray}
with a constant $C_{\rm deloc}$. To derive Eq.~\eqref{devi_stat4_hosoku}, we use the fact that the number of the set $\mathcal{D}$ is proportional to $\sqrt{M}$.
In the large system-size limit, we obtain
\begin{eqnarray}
\triangle x _{\rm stat} \propto M^{3/4}. 
\label{devi_stat5}
\end{eqnarray}

This behavior is anomalous because the deviation is not proportional to the system size $M$. 
In a non-interacting fermion model without disorder, the particle extends to the whole system, and thus the deviation scales as $M$.
From the same reason as above, the deviation in the AAM with $W<1$ also scales as $M$.
In stark contrast to the AAM, the RDM does obey Eq.~\eqref{devi_stat5} owing to the existence of many LESs even in the delocalized phase.

\section{Experimental possibility}
The Anderson localization in 1D systems has been observed using a quasi-periodic potential \cite{AL_gas1} and random speckle potentials \cite{AL_gas2,AL_gas3,AL_gas4,AL_gas5,AL_gas6,AL_gas7} in cold atoms. The former case corresponds to the AAM, and our prediction can be accessible by using the quantum gas microscope. On the other hand, the RDM can in principle be realized using digital micromirror devices, by which various kinds of potentials including a random one have already been made in a highly controllable manner \cite{DMD1,DMD2,DMD3}.

\end{document}